# Reversible spin-optical interface in luminescent organic radicals

Sebastian Gorgon[1,2,*], Kuo Lv[3], Jeannine Grüne[4], Bluebell H. Drummond[1], William K. Myers[2], Giacomo Londi[5], Gaetano Ricci[5], Danillo Valverde[5], Claire Tonnelé[6], Petri Murto[7], Alexander S. Romanov[8], David Casanova[6], Vladimir Dyakonov[4], Andreas Sperlich[4], David Beljonne[9], Yoann Olivier[5], Feng Li[3], Richard H. Friend[1,*], Emrys W. Evans[10,*]

[1] Cavendish Laboratory, University of Cambridge; JJ Thomson Ave, Cambridge, CB3 0HE, United Kingdom
[2] Centre for Advanced Electron Spin Resonance, Department of Chemistry, University of Oxford; Inorganic Chemistry Laboratory, S Parks Rd, Oxford, OX1 3QR, United Kingdom
[3] State Key Laboratory of Supramolecular Structure and Materials, College of Chemistry, Jilin University; Qianjin Avenue 2699, Changchun, 130012, P. R. China
[4] Experimental Physics VI, Faculty of Physics and Astronomy, University of Würzburg; 97074 Würzburg, Germany
[5] Laboratory for Computational Modelling of Functional Materials, Namur Institute of Structured Matter, University of Namur; Rue de Bruxelles 61, 5000 Namur, Belgium
[6] Donostia International Physics Centre; Donostia, Euskadi, Spain
[7] Yusuf Hamied Department of Chemistry, University of Cambridge; Cambridge, CB2 1EW, United Kingdom
[8] Department of Chemistry, University of Manchester; Manchester, M13 9PL, United Kingdom
[9] Laboratory for Chemistry of Novel Materials, University of Mons; 7000 Mons, Belgium
[10] Department of Chemistry, Swansea University; Singleton Park, Swansea, SA2 8PP, United Kingdom

*Corresponding authors. Email addresses: sg911@cam.ac.uk (S.G.), rhf10@cam.ac.uk (R.H.F.), emrys.evans@swansea.ac.uk (E.W.E.).

**Molecules present a versatile platform for quantum information science,[1,2] and are candidates for sensing and computation applications.[3,4] Robust spin-optical interfaces are key to harnessing the quantum resources of materials.[5] To date, carbon-based candidates have been non-luminescent,[6,7] which prevents optical read-out. Here we report the first organic molecules displaying both efficient luminescence and near-unity generation yield of high-spin multiplicity excited states. This is achieved by designing an energy resonance between emissive doublet and triplet levels, here on covalently coupled tris(2,4,6-trichlorophenyl) methyl-carbazole radicals (TTM-1Cz) and anthracene. We observe the doublet photoexcitation delocalise onto the linked acene within a few picoseconds and subsequently evolve to a pure high spin state (quartet for monoradicals, quintet for biradical) of mixed radical-triplet character near 1.8 eV. These high-spin states are coherently addressable with microwaves even at 295 K, with optical read-out enabled by intersystem crossing to emissive states. Furthermore, for the biradical, on return to the ground state the previously uncorrelated radical spins either side of the anthracene show strong spin correlation. Our approach simultaneously supports a high efficiency of initialisation, spin manipulations and light-based read-out at room temperature. The integration of luminescence and high-spin states creates an organic materials platform for emerging quantum technologies.**

Considerable progress has been made towards designing molecular systems fulfilling the DiVincenzo criteria for practical qubits.[8] Optical addressability has been demonstrated in organometallic complexes with triplet ground states at liquid helium temperatures.[9] Related complexes display impressive spin coherence times, reaching the microsecond range at room temperature.[10] Structures without metal atoms can be more immune to decoherence,[11] and such fully organic molecules have been used in several demonstrations of quantum effects.[12–14]

Radical organic molecules contain unpaired electrons which can be stabilised by chemical design. Advances have been made using non-luminescent radicals that are covalently attached to chromophores, and such structures can support high spin multiplicity ($S > 1$) excited states.[6,7] The presence of a radical can enhance the rate of intersystem crossing (ISC), leading to accumulation of chromophore triplet states.[15] If exchange between the triplet ($S = 1$) and radical ($S = 1/2$) spin is larger than all other magnetic interactions, a distinct quartet ($S = 3/2$) and doublet ($S = 1/2$) state pair forms.[16] If a second radical is additionally coupled, a quintet ($S = 2$) state can be achieved.[17] High-spin states allow building dense architectures with multiple qubits hosted within a single manifold of spin sublevels.[18] Such multilevel qubits, termed qudits, offer scalability advantages in quantum computation.[19] The qudit behaviour of quartet states was recently demonstrated in PDI-TEMPO at 80 K.[20] However, current high-spin structures have large (~1 eV) energy gaps between the photogenerated chromophore singlet state and the triplet manifold.[6,17,21] Critically, this prevents reverse intersystem crossing (RISC) to a luminescent state. Thus, all organic high-spin systems to date are non-emissive, which makes optical read-out impossible.

While most stable radicals are non-emissive, there is now a class of luminescent radicals that offer fully spin-allowed emission within the doublet manifold.[22] The set of available molecular structures is rapidly expanding and spans a broad wavelength range.[23,24] Record efficiencies for deep-red and infrared light emitting diodes (LEDs) were recently reached in tris(2,4,6-trichlorophenyl) methyl (TTM) radicals linked to carbazole electron donors.[25]

By utilising a doublet ($D_1$) level with substantial oscillator strength for absorption and emission, we can avoid excitation via the singlet state in radical-chromophore structures. In our designs, we bring the triplet ($T_1$) level into energy resonance with a $D_1$ level on a luminescent radical. Eliminating the energy gap between the photogenerated and high-spin states allows interconversion in either direction. This makes it possible to initialise and, for the first time, optically read-out high-spin states in organic molecules.

## Optical properties

We use TTM-1Cz as a luminescent radical unit "R" as it displays 41% photoluminescence quantum efficiency (PLQE) for red emission in dilute toluene solutions (Fig. 1a).[26] Its small size allows proximity between the radical and chromophore. To achieve doublet-triplet energy resonance involving the emissive excited state with $E(D_1) = 1.82$ eV, we select anthracene with $E(T_1) = 1.83$ eV (Fig. 1b), linked at its 9-position to the para-position of the TTM-1Cz carbazole, thus preserving through-bond conjugation (SI Section 1). We prepare two "R-A" monoradical structures (TTM-1Cz-An, and TTM-1Cz-PhAn with a bridging phenyl ring), and a "R-A-R" biradical $(TTM\text{-}1Cz)_2$-An. At room temperature their absorption spectra are only weakly modified, and their photoluminescence (PL) shows a red-shift relative to TTM-1Cz (Fig. 1c). The PLQE in toluene solution is 32% for TTM-1Cz-An, 4% for TTM-1Cz-PhAn and 3% for $(TTM\text{-}1Cz)_2$-An.

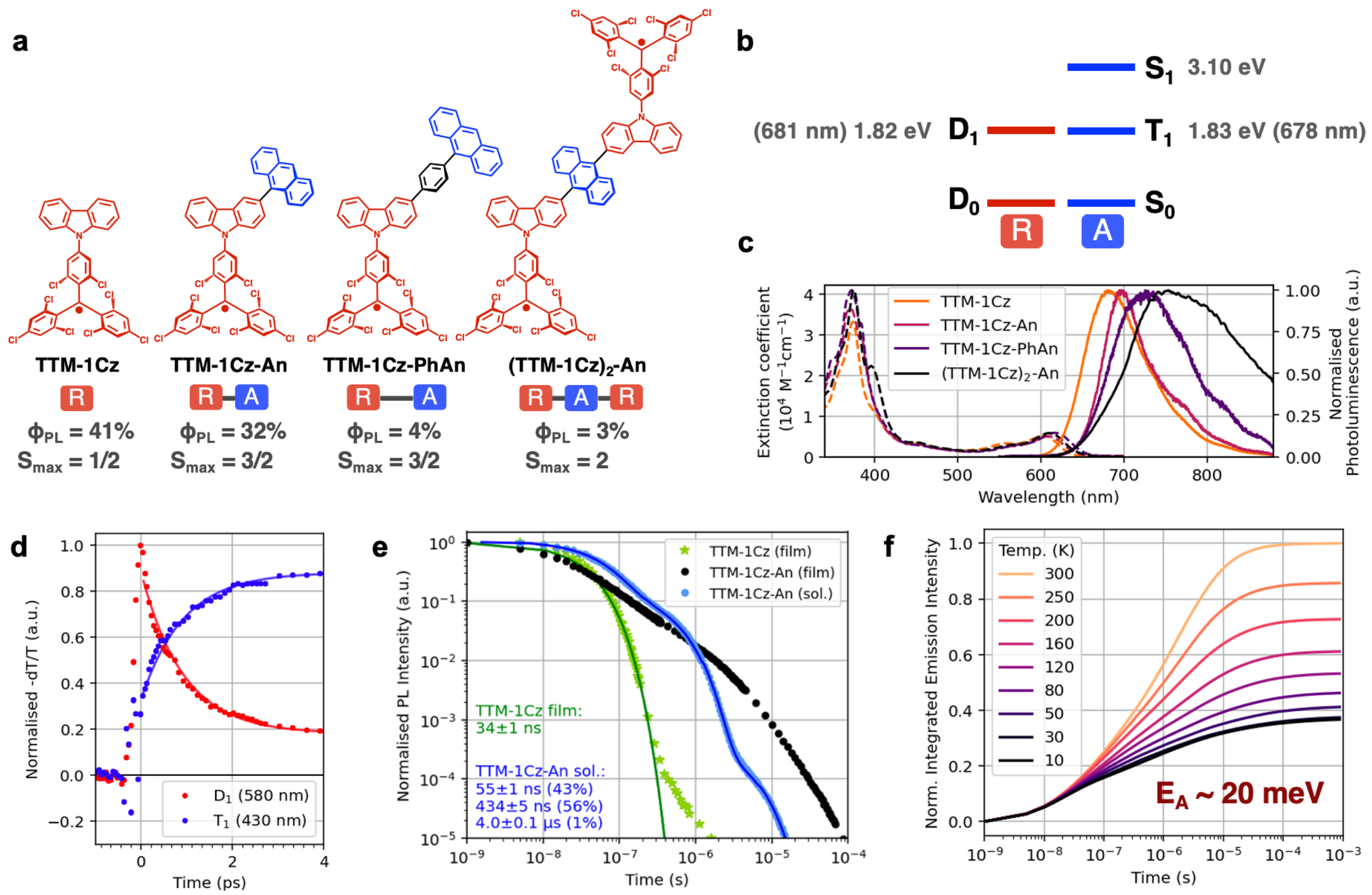

**Fig. 1. Luminescent radical-acene molecular system. a,** Molecules featured in this study, their dilute toluene solution PLQE ($\phi_{PL}$) and the highest spin quantum number of their excited state ($S_{max}$). **b,** Energy levels for TTM-1Cz and anthracene, extracted from emission data in separated molecules, revealing the $D_1$-$T_1$ energy alignment. **c,** Steady-state absorption and emission spectra obtained after 532 nm excitation in 200 μM toluene solutions at room temperature. Anthracene-linked compounds show a small redshift relative to TTM-1Cz. **d,** Ultrafast transient absorption kinetics of dilute toluene TTM-1Cz-An solution following a 600 nm pulse at 295 K extracted from photoinduced absorption features. Solid lines show 0.92 ps lifetime fits. **e,** Emission kinetics of 5% in PMMA films and solutions of TTM-1Cz and TTM-1Cz-An at 295 K following 532 nm excitation. **f,** Temperature dependence of integrated emission intensity of 5% TTM-1Cz-An in PMMA films after 520 nm excitation revealing that the radical emission is temperature activated.

We perform transient optical absorption (TA) spectroscopy to establish the mechanism of interaction between the radical and linked anthracene. We selectively excite the radical using a 600 nm pulse, well below the anthracene singlet absorption onset (400 nm). In dilute solutions of R-A and R-A-R we observe rapid decay of the $D_1$ photoinduced absorption (PIA) and a matching rise of the $T_1$ PIA, indicating transfer of the excitation from the TTM-1Cz to the anthracene with local $T_1$ character (Ext. Data Fig. 1). This occurs with an ultrafast lifetime of 0.9±0.2 ps in TTM-1Cz-An (Fig. 1d), 5.6±0.6 ps in TTM-1Cz-PhAn, and 0.7±0.2 ps in (TTM-1Cz)$_2$-An (Fig. S12).

We focus on TTM-1Cz-An dynamics as it is the most emissive material. The prompt TA spectra reveal that partial local triplet character is present already within our 150 fs time resolution. When the solvent polarity is increased from cyclohexane to toluene, there are no changes to the rapid energy transfer dynamics, and we extract a 93% yield of $T_1$ generation (Fig. S13). Subsequently, a new PIA near 685 nm appears around ~30 ps and we assign it to an intramolecular charge transfer (CT) state as its spectral position matches the anthracene radical cation.[27] The population of the CT peaks around 100 ps in toluene solutions. In more polar 2-methyltetrahydrofuran (MeTHF) we observe complete non-radiative decay within 1 ns, likely via a low-lying CT state.

We perform time-resolved emission spectroscopy to understand how $D_1$ emission is preserved despite rapid energy transfer dynamics. TTM-1Cz-An emission shows triexponential kinetics in toluene solutions at room temperature (Fig. 1e). The emission lineshape is unchanged throughout the decay (Fig. S14). The fastest emission component (55±1 ns) is two-fold slower compared to the monoexponential lifetime of TTM-1Cz in the same solvent (27±1 ns). Most of TTM-1Cz-An emission is delayed even further and occurs with a lifetime of 434±5 ns. Since the $D_1$ character is lost with near unity yield well before the TTM-1Cz emission lifetime, all radical emission in TTM-1Cz-An is preceded by a temperature-activated process. TTM-1Cz-An emission lineshape becomes more structured as the temperature is lowered and resembles anthracene phosphorescence at 10 K (Fig. S15). A vibronic progression is present in both excitation and emission scans at 77 K, similar to that of anthracene triplets (Ext. Data Fig. 2).[28] These features could signify a partial $T_1$-like character of the photogenerated state.

We perform temperature-dependent time-resolved emission spectroscopy in dilute poly(methyl methacrylate) (PMMA) films (Ext. Data Fig. 3). We find that the total emission intensity strongly increases with temperature (Fig. 1f), in contrast to the temperature-independent intensity and rate of $D_1$ emission in TTM-1Cz (Fig. S16). Using an Arrhenius-type model, we estimate the activation energy for emission in TTM-1Cz-An to be approx. 20 meV. All R-A and R-A-R display temperature-activated emission and structured phosphorescence at low temperatures (Fig. S17). The activation energy for delayed emission in the $(TTM\text{-}1Cz)_2$-An biradical is approx. 12 meV.

## Spin properties

Our optical experiments show rapid generation of a local triplet character excited state followed by temperature activated emission with $D_1$ character. We use ESR to probe the spin properties of the states involved in this mechanism. Continuous wave (cw) ESR in the dark shows a narrow signal centred at $g = 2.0036$ for all four molecules studied (Ext. Data Fig. 4), characteristic of a TTM $D_0$ transition. We perform transient cw ESR (trESR) to track the excited state sublevel dynamics following selective $D_1$ excitation with a pulsed laser.[29] At X-band in the half field (HF) region, R-A show signals centred at $g = 4.04$ and $g = 6.20$ (Fig. 2a). These are first-order forbidden transitions with $\Delta m_s = 2$ and $\Delta m_s = 3$ respectively, where the latter gives clear evidence of probing a quartet state (Fig. S18). In the full field (FF) region, dilute frozen toluene solutions of R-A show a broad $\Delta m_s = 1$ signal with a width of ~103 mT, superimposed with a narrow signal at $g = 2.00$ (Fig. 2b). The signal width is reduced by a factor of 2/3 compared to one expected from anthracene triplets (~152 mT).[30] This suggests that a quartet state is formed due to strong exchange between the anthracene triplet-like wavefunction and the ground state radical spin. Absence of level crossings in trESR at Q-band indicates that the R-A quartet-doublet energy gap is at least 0.8 meV (Ext. Data Fig. 5).

Using a transient nutation pulse sequence, we can directly confirm the spin multiplicity of the sublevels involved in ESR transitions.[31] In the dark, only the doublet nutation frequency ($\omega_0$) is detected for both R-A. Following 600 nm light excitation, we find that there is no contribution from triplet transitions ($\omega = \sqrt{2}\omega_0$) across the entire spectra of R-A. The broad feature is due to "outer" quartet transitions ($\omega = \sqrt{3}\omega_0$), while the narrow central features are due to both "inner" quartet ($\omega = 2\omega_0$) and doublet ($\omega = \omega_0$) transitions (Fig. 2c). This confirms the strong exchange regime holds throughout the entire molecular ensemble, regardless of conformational effects.

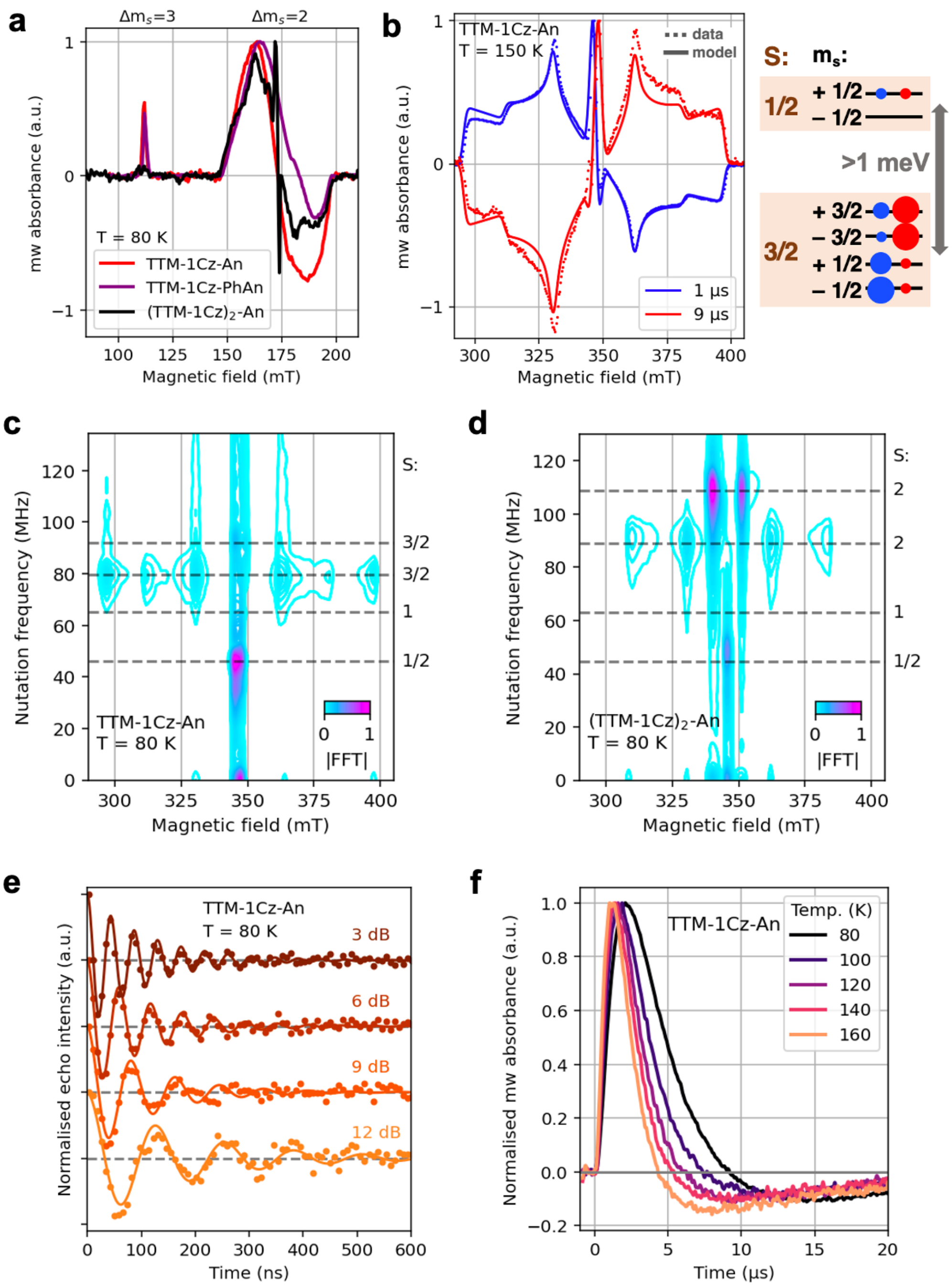

**Fig. 2. ESR on high-spin states. a,** Transient half field cw X-band ESR spectra at 80 K following 600 nm excitation showing $\Delta m_s = 3$ and $\Delta m_s = 2$ transitions in frozen toluene solutions of R-A, and a patterned $\Delta m_s = 2$ in R-A-R. **b,** Transient full field cw X-band ESR $\Delta m_s = 1$ spectra of TTM-1Cz-An frozen toluene solution at 150 K following 600 nm excitation and schematic population patterns extracted from the model. Size of circles is proportional to the sublevel population at early (blue) and late (red) times. **c,** Transient nutation at 80 K of dilute frozen toluene solution of TTM-1Cz-An following 600 nm excitation indicating quartet ($S = 3/2$) state formation. **d,** Transient nutation at 80 K of dilute frozen toluene solution of (TTM-1Cz)$_2$-An following 600 nm excitation revealing quintet ($S = 2$) multiplicity of the polarized signal. **e,** Rabi oscillations on the TTM-1Cz-An quartet (331.5 mT) at 80 K as a function of microwave power. **f,** Temperature dependence of TTM-1Cz-An quartet (331.5 mT) polarisation inversion.

The presence of strong exchange in R-A indicates the anthracene-carbazole linkage supports significant wavefunction delocalisation. In the symmetrically substituted $(TTM\text{-}1Cz)_2$-An biradical the FF trESR spectrum is further narrowed (Fig. S19). Its width of ~77 mT is consistent with formation of a quintet state in the strong exchange regime. The polarisation pattern reveals an ISC population mechanism, as in R-A quartets. Transient nutation confirms that the broad spectrum of $(TTM\text{-}1Cz)_2$-An is due to quintet transitions ($\omega = \sqrt{6}\omega_0$ or $\omega = 2\omega_0$), while the narrow central feature is exclusively due to a doublet transition (Fig. 2d). We do not detect any quartet or uncoupled triplet features, which shows that the triplet-character excitation on the anthracene in R-A-R is strongly coupled to both radical electrons within an overall four-spin state.

Turning to the most luminescent TTM-1Cz-An, we have explored Rabi oscillation experiments to quantify their potential as molecular qudits coupled to emission.[32] We place the quartet in a coherent superposition which is then probed with a Hahn echo sequence (Fig. 2e). This allows us to find the quantum fidelity $\Omega_M = 2T_m\omega_R$, where $T_m$ is the spin coherence time, and $\omega_R$ is the Rabi frequency. At 80 K, $T_m$ is 1.5±0.1 µs, and $\Omega_M$ values up to ~70 were found, comparable to non-luminescent high-spin states.[20] The $\Omega_M$ values scale linearly with microwave power, showing that this quartet state can be placed into an arbitrary superposition.

We model the R-A trESR spectra to track the sublevel population dynamics (Ext. Data Tab. 1). The prompt signal shows a majority population of the quartet $m_s$ = ±1/2 sublevels at all temperatures (Fig. 2b diagram). The polarisation pattern inverts at later times, with the inversion occurring faster upon increasing temperature (Fig. 2f). By fitting the quartet polarisation inversion times to an Arrhenius-type model, we find an activation energy of 19±1 meV for TTM-1Cz-An (Ext. Data Fig. 5). This is in excellent agreement with the activation energy for emission found by optical spectroscopy and confirms that $D_1$ emission is preceded by RISC from the quartet state. The $m_s$ = ±3/2 sublevels form the majority of the quartet population after inversion, consistent with a preferential depopulation of the $m_s$ = ±1/2 sublevels during RISC.

**Luminescent R-A mechanism**

Our ESR results show that a complete description of the electronic states in R-A and R-A-R requires knowledge of their total spin multiplicity as well as of the nature of the contributing anthracene-like and TTM-1Cz-like states (Fig. 3a). Multi-Configurational Self-Consistent Field (MCSCF) calculations for TTM-1Cz-An (SI Section 3) show the photogenerated state as a transition between the $H_R$ and $S_R$ molecular orbitals (MOs). This state is similar to the $D_1$ state in TTM-1Cz, but additionally contains some wavefunction density on the linked anthracene in the $H_R$ MO (Fig. 3b). At the ground-state equilibrium geometry, this state contains approx. 5% contribution on the anthracene (Fig. S26), and lies 13 meV above states with local triplet character (Fig. S22, Tab. S2). The calculations also reveal the presence of an intramolecular charge-transfer state ($^2$CT) from the anthracene to the TTM-1Cz moiety that spans a broad energy range from ~60 meV above the photogenerated state at the equilibrium geometry to ~20 meV below that state for an orthogonal arrangement of the anthracene unit (Tab. S5). We have developed a kinetic model for the energy transfer which includes a $^2$CT intermediate (Tab. S15). We compute energy transfer times spanning a 0.1-10 ps range when close to the ground-state equilibrium conformation. The calculated quartet state spin Hamiltonian parameters agree well with those extracted from modelling the ESR data (Fig. S25, Tab. S9).

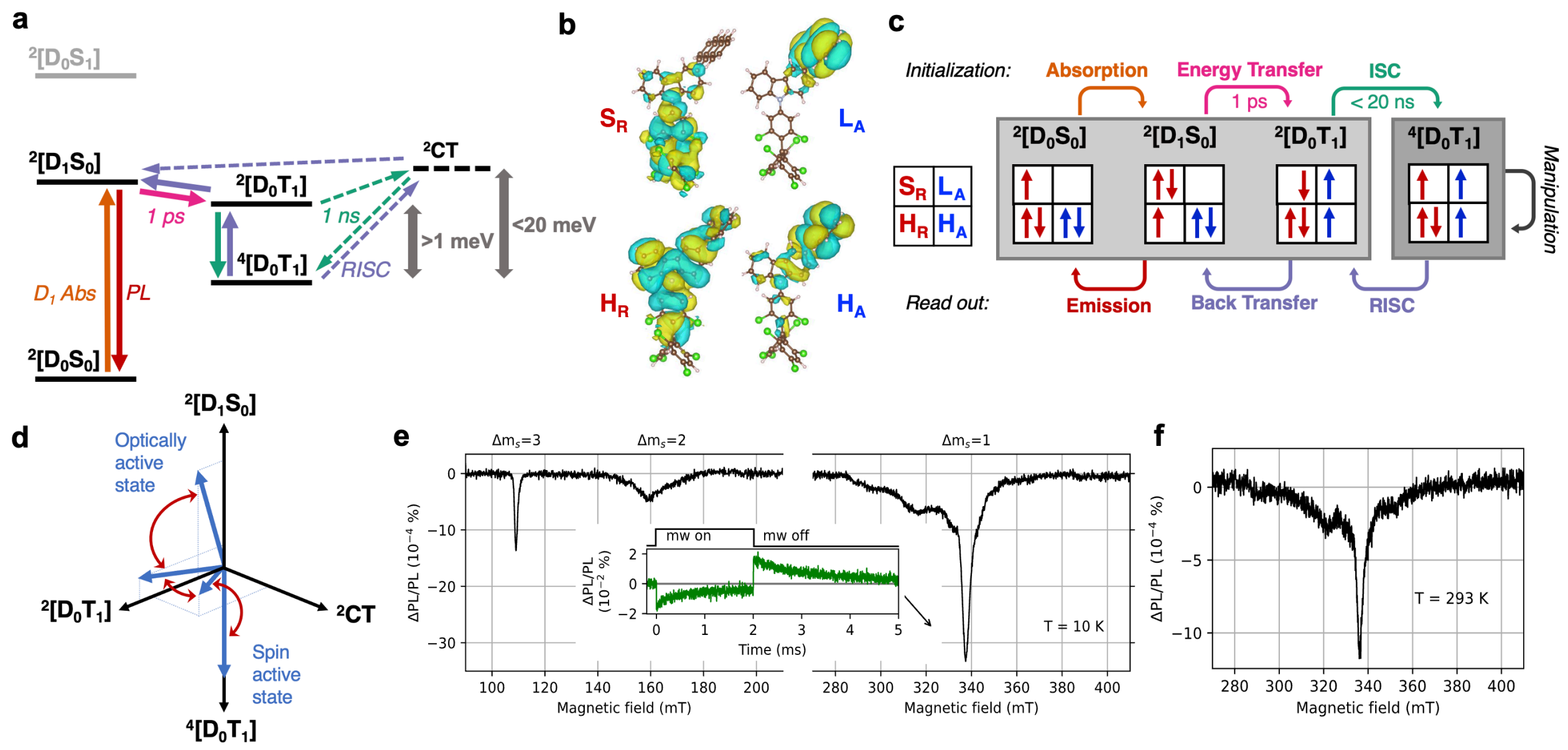

**Fig. 3. Luminescent R-A mechanism and optical read-out. a,** R-A energy level diagram showing rapid quartet state generation following light absorption by the radical. At room temperature, activation to the $^2[D_1S_0]$ state is efficient and only radical photoluminescence is seen. **b,** Molecular orbitals of TTM-1Cz-An obtained with MCSCF calculations ($H_R$: Radical HOMO, $S_R$: Radical SOMO, $H_A$: Acene HOMO, $L_A$: Acene LUMO). **c,** Summary scheme of dominant orbital contributions during quartet formation and radical emission. **d,** Vectorial depiction of the evolution of state mixing during reversible quartet formation. Character of the photoexcited doublet state evolves from majority $^2[D_1S_0]$ to majority $^2[D_0T_1]$ on ultrafast timescales. If energetically accessible, the $^2$CT state may be additionally involved in the doublet state mixing and thus assist the ISC to a pure quartet state. **e,** ODMR spectra of dilute frozen toluene solution of TTM-1Cz-PhAn under 532 nm excitation at 10 K, showing the participation of the quartet state in the FF and HF region. Inset: trODMR at 337.5 mT showing a PL reduction in resonance. **f,** ODMR spectra of dilute PMMA films of TTM-1Cz-PhAn under 532 nm excitation at 293 K.

In the pure diabatic eigenstate description of R-A, absorption of light with energies below singlet anthracene band gap occurs from $^2[D_0S_0]$ to $^2[D_1S_0]$ (Fig. 3c). This photoexcited state is energetically close to $^2[D_0T_1]$ and rapid localisation of the wavefunction onto the coupled anthracene occurs within a few ps. Such high rates are possible as the total spin multiplicity is conserved during $^2[D_1S_0]$ to $^2[D_0T_1]$ energy transfer. Subsequently, the $^2[D_0T_1]$ exciton undergoes ISC to the $^4[D_0T_1]$ quartet state. Since the spatial wavefunctions of the $^2[D_0T_1]$ and $^4[D_0T_1]$ levels are nearly identical, the direct SOC matrix element is small (Tab. S6).[33] Direct ISC is thus unlikely to generate the high quartet yields on sub-20 ns timescales as implied by the observed emission dynamics. We consider that the $^2$CT state is close in energy to the $^{2,4}[D_0T_1]$ excitonic states. This is supported by solvatochromism observed in TA and MCSCF calculations. Accessibility of a $^2$CT state might assist ISC to the quartet state via SOC, which is larger for states of different character (Tab. S8). The $^2$CT state can form from $^2[D_0T_1]$ following spin-conserving electron transfer from $L_A$ onto $S_R$. If a spin flip occurs during the back-transfer, the $^4[D_0T_1]$ state forms. This forward mechanism is near-barrierless and supports our observations of a high yield of quartet states at temperatures from 20 to 300 K. We can estimate the quartet state yield using luminescence dynamics and PLQE, together with the $T_1$ yield extracted from TA. The approximate yield of $^4[D_0T_1]$ states in toluene solutions of TTM-1Cz-An at room temperature is ~73%.

The process preceding emission is temperature-activated. Due to low energetic barriers present in our molecular design, only ~22% of the total emission yield is lost in TTM-1Cz-An compared to TTM-1Cz at room temperature. Reforming a spin doublet state from the quartet is the rate-limiting step due to the need for a change in spin multiplicity. We assign the ~20 meV activation energy extracted from optical and trESR spectroscopy to $^4[D_0T_1] \rightarrow {}^2CT$ transfer. After the $^2CT$ state is reformed, a hole transfer from anthracene onto the carbazole can rapidly yield the emissive $^2[D_1S_0]$ state. At temperatures below 100 K, phosphorescence is observed. This can originate either from the $^4[D_0T_1]$ in a spin-forbidden process at the lowest temperatures, or from the $^2[D_0T_1]$ state in a spin-allowed process. The increase in phosphorescence intensity with temperature in this regime confirms that $^4[D_0T_1]$ is the lowest energy excited state (i.e., that the exchange is ferromagnetic).

The presence of a bridging phenyl in TTM-1Cz-PhAn leads to a large loss of PLQE, as both energy transfer (Fig. S12), RISC and emission (Ext. Data Fig. 3) are slower compared to TTM-1Cz-An. The observed PL redshift could be due to a smaller electron-hole overlap in TTM-1Cz-PhAn, which leads to a less emissive state. Despite the greater spatial separation between the radical and anthracene, TTM-1Cz-PhAn remains in the strong exchange coupling regime (Ext. Data Fig. 5).

The small energy offsets between excited states involved in this mechanism likely lead to a high degree of mixing between the three states with overall doublet multiplicity (Fig. 3d), as indicated by calculated electronic couplings (Tab. S12).[34] We detect signatures of $^2[D_1S_0]$ and $^2[D_0T_1]$ mixing in our ultrafast optical spectroscopy and in low temperature excitation scans. While these effects are likely modulated by conformational reorganisation, vibrational motion and environment dynamics, the quartet-doublet energy gap is always in the strong exchange regime, as demonstrated by our ESR experiments. Therefore, our system benefits from low energy offsets on the optical scales, but large energy offsets on the magnetic scales. This enables access to robust high-spin excited states coupled to an efficient emissive state.

**Optical read-out at room temperature**

The luminescence of our materials opens the path toward optical read out in organic high-spin molecules, which we explore with optically detected magnetic resonance (ODMR) measurements. Dilute frozen toluene solutions of R-A at 10 K show a broad, patterned FF ODMR signal (Fig. 3e), matching quartet simulation parameters of trESR. As in trESR, we also observe HF ODMR signals with $\Delta m_s = 3$ and $\Delta m_s = 2$ transitions at $g = 6.17$ and $g = 4.23$ respectively. To achieve read-out at ambient temperatures we perform ODMR on dilute PMMA films of the two R-A molecules. Both TTM-1Cz-An and TTM-1Cz-PhAn films show a clear ODMR contrast at room temperature. The resonant PL change in films at 295 K is comparable to that observed at 10 K (Ext. Data Fig. 6). The dipolar parameters remain unchanged, confirming assignment to the quartet state. No ODMR signals were detected for TTM-1Cz, indicating identical emissive rates for the $m_s = \pm 1/2$ $D_1$ sublevels in an isolated radical. The ODMR contrast in R-A is thus due to $^4[D_0T_1]$ origin of $^2[D_1S_0]$ emission, and demonstrates optical readout of the quartet state at room temperature (Fig. 3f).

To investigate the sign of the ODMR signal, we perform transient ODMR (trODMR) measurements, which directly probe the PL change in resonant conditions using digitizer detection.[35] ODMR transients with application of 2 ms square mw pulse at both FF and HF reveal a negative sign of all signals (Fig. 3e inset). While at lower temperatures the quartet state is coupled to phosphorescence, at higher temperatures quartet depopulation occurs via RISC. The negative sign of the ODMR signal suggests that microwaves drive transitions from the more populated $m_s$ = ±1/2 to less efficiently linked $m_s$ = ±3/2 quartet sublevels, thus decreasing the PL from the doublet state in resonant conditions.

As in ODMR, we observe spin polarised quartet signals in room temperature trESR on PMMA films of R-A (Fig. 4a). We additionally perform room temperature pulsed ESR experiments (Ext. Data Fig. 7). Quartet echoes were detectable up to 10 μs after photoexcitation at 295 K, further demonstrating the potential of these molecules as optically addressable qudits in the solid state.

**Biradical ground state control**

The presence of two radical spins in luminescent R-A-R allows engineering of more complex ground and excited state interactions. As shown above, the photophysical properties of the (TTM-1Cz)$_2$-An biradical are analogous to TTM-1Cz-An, with ultrafast wavefunction localisation onto the anthracene after light absorption followed by delayed emission. However, the biradical nature of R-A-R has significant consequences for the spin properties compared to R-A monoradicals. In the R-A-R ground state the exchange interaction between the two $D_0$ electrons is extremely weak (≤μeV) as revealed by the exclusively doublet transient nutation signals in the dark and temperature dependence of cw ESR intensity (Fig. S20). This is consistent with the large spin-spin distance of 2.09±0.19 nm determined by double electron-electron resonance (DEER) spectroscopy (Fig. S21). Therefore, the biradical spin pair is uncorrelated in the ground state, and the $^1[D_0S_0D_0]$ and $^3[D_0S_0D_0]$ states are degenerate, as in a true biradical.[36]

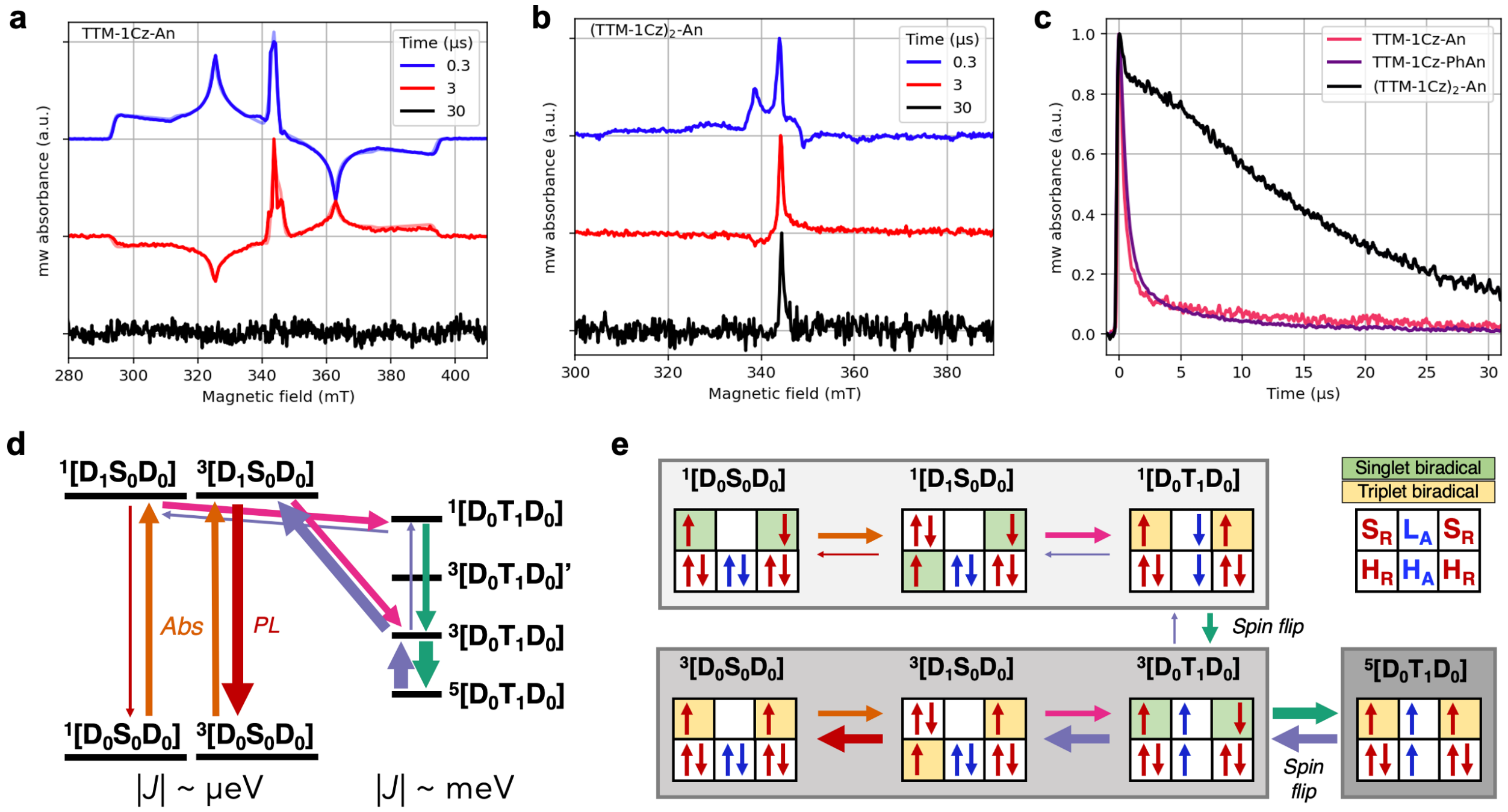

**Fig. 4. Room temperature spin dynamics and ground state control. a,** TrESR spectra on 5% PMMA films of TTM-1Cz-An under 600 nm excitation at 295 K (solid line) and their simulation (translucent lines). Noise of 30 μs trace scaled to match noise level in panel b. **b,** TrESR spectra on 5% PMMA films of (TTM-1Cz)$_2$-An under 532 nm excitation at 295 K showing a long lived $g$ = 2.00 (344.5 mT) signal. **c,** Comparison of $g$ = 2.00 kinetics in trESR on 5% PMMA films at 295 K. Long lived doublet polarisation is seen in R-A-R but not in R-A. **d,** (TTM-1Cz)$_2$-An energy level diagram showing two exchange coupling regimes. In the ground and initially photogenerated states the radical spins are uncorrelated (|$J$| ~ μeV), but become strongly coupled when the $T_1$ wavefunction is present (|$J$| ~ meV) **e,** Summary scheme of dominant orbital contributions during quintet formation. The triplet channel is more kinetically favoured during temperature activated emission, which leads to a strong polarisation of the ground state.

Coupling between $D_0$ electrons can be transiently switched on due to strong (~meV) exchange within $[D_0T_1D_0]$ excited states, and we detect polarised $^5[D_0T_1D_0]$ in room temperature trESR (Fig. 4b). The absence of $^3[D_0T_1D_0]$ triplet signals confirms that the quintet is the lowest excited state and is generated efficiently. Following relaxation back to the ground state, the spin information of the $D_0$ biradical electrons is preserved. This is evidenced by the long-lived $g$ = 2.00 polarisation, extending beyond 30 μs at room temperature (Fig. 4c), well past any luminescence or quintet polarisation. Such ground state polarisation is absent in R-A monoradicals.

To understand its origin, we examine the energy level structure of R-A-R. As the singlet and triplet ground states are degenerate, both contribute to the forward process of quintet state generation (Fig. 4d). The singlet and triplet channels each provide an efficient, rapid, spin-conserving pathway to a $[D_0T_1D_0]$ state, analogous to the doublet pathway in R-A. However, as the quintet depopulation involves one spin-flip to access $^3[D_0T_1D_0]$ but two spin-flips to reach $^1[D_0T_1D_0]$, the triplet channel dominates the reverse process (Fig. 4e). Since the $^3[D_1S_0D_0]$ state can then quickly form and decay radiatively, this leads to an excess of $^3[D_0S_0D_0]$ ground state. Despite the ground state energetic degeneracy, this method allows for a preferential preparation of the triplet biradical configuration. Only following light-induced quintet generation and relaxation, the ground state biradical consists of a spin-polarized pair of electrons which loses spin alignment with a long lifetime of 20±1 μs at room temperature in solid films. This timescale matches spin-lattice relaxation time of $D_0$ electrons measured in pulsed ESR on the same $(TTM\text{-}1Cz)_2$-An film at 295 K (Ext. Data Fig. 7), and could be extended further by host engineering and deuteration.

**Conclusions**

Through our study combining optical spectroscopy, ESR, ODMR and theoretical modelling, we have demonstrated that we can generate pure high-spin states in organic molecules, manipulate them, and then read them optically (Fig. 3c). By using a luminescent radical and engineering an excited state manifold with small energy offsets, we have revealed a new pathway to generate high-spin states, here exemplified as a quartet or quintet. The radical can dress the triplet exciton in hybridised states, which can reversibly access the high-spin manifold. Consequently, a luminescent state can be restored from addressable high-spin states. Remarkably, we observe polarised high-spin states via ESR and ODMR at room temperature in non-crystalline solid state, showing real potential for future applications such as quantum sensing.

Our approach of coherently manipulating high-spin states coupled to emission complements insights gained from alternative qubit platforms, such as colour centres, in particular the diamond nitrogen-vacancies (NV), where optical spin polarisation is also generated by excited-state intersystem crossing. While $NV^-$ centres display long spin coherence times at room temperature, their scalability may be limited by challenges of controlling defect placement and preventing decoherence in non-isolated defects.[37] In contrast, molecules offer unparalleled chemical tunability due to a wealth of synthetic approaches, and potential to develop more extended spin structures, as demonstrated here by the behaviour of R-A and R-A-R. The scope for chemical tuning and for extension to polyradical structures opens new opportunities for designs of coupled spin systems which can be addressed with light in target wavelengths ranges, noting that their placement and intermolecular interactions can be controlled with self-assembly and scanning probe methods.

## References

1. Wasielewski, M. R. *et al.* Exploiting chemistry and molecular systems for quantum information science. *Nat. Rev. Chem.* **4**, 490–504 (2020).
2. Atzori, M. & Sessoli, R. The Second Quantum Revolution: Role and Challenges of Molecular Chemistry. *J. Am. Chem. Soc.* **141**, 11339–11352 (2019).
3. Yu, C.-J., von Kugelgen, S., Laorenza, D. W. & Freedman, D. E. A Molecular Approach to Quantum Sensing. *ACS Cent. Sci.* **7**, 712–723 (2021).
4. Gaita-Ariño, A., Luis, F., Hill, S. & Coronado, E. Molecular spins for quantum computation. *Nat. Chem.* **11**, 301–309 (2019).
5. Awschalom, D. D., Hanson, R., Wrachtrup, J. & Zhou, B. B. Quantum technologies with optically interfaced solid-state spins. *Nat. Photonics* **12**, 516–527 (2018).
6. Quintes, T., Mayländer, M. & Richert, S. Properties and applications of photoexcited chromophore–radical systems. *Nat. Rev. Chem.* (2023) doi:10.1038/s41570-022-00453-y.
7. Teki, Y. Excited-State Dynamics of Non-Luminescent and Luminescent π-Radicals. *Chem. - A Eur. J.* **26**, 980–996 (2020).
8. DiVincenzo, D. P. The Physical Implementation of Quantum Computation. *Fortschritte der Phys.* **48**, 771–783 (2000).
9. Bayliss, S. L. *et al.* Optically addressable molecular spins for quantum information processing. *Science (80-. ).* **370**, 1309–1312 (2020).
10. Bader, K. *et al.* Room temperature quantum coherence in a potential molecular qubit. *Nat. Commun.* **5**, 1–5 (2014).
11. Warner, M. *et al.* Potential for spin-based information processing in a thin-film molecular semiconductor. *Nature* **503**, 504–508 (2013).
12. Wang, D. *et al.* Turning a molecule into a coherent two-level quantum system. *Nat. Phys.* **15**, 483–489 (2019).
13. Filidou, V. *et al.* Ultrafast entangling gates between nuclear spins using photoexcited triplet states. *Nat. Phys.* **8**, 596–600 (2012).
14. Oxborrow, M., Breeze, J. D. & Alford, N. M. Room-temperature solid-state maser. *Nature* **488**, 353–356 (2012).
15. Giacobbe, E. M. *et al.* Ultrafast Intersystem Crossing and Spin Dynamics of Photoexcited Perylene-3,4:9,10-bis(dicarboximide) Covalently Linked to a Nitroxide Radical at Fixed Distances. *J. Am. Chem. Soc.* **131**, 3700–3712 (2009).
16. Kollmar, C. & Sixl, H. Theory of a coupled doublet-triplet system. *Mol. Phys.* **45**, 1199–1208 (1982).
17. Teki, Y., Miyamoto, S., Nakatsuji, M. & Miura, Y. π-topology and spin alignment utilizing the excited molecular field: Observation of the excited high-spin quartet (S = 3/2) and quintet (S = 2) states on purely organic π-conjugated spin systems. *J. Am. Chem. Soc.* **123**, 294–305 (2001).
18. Leuenberger, M. N. & Loss, D. Quantum computing in molecular magnets. *Nature* **410**, 789–793 (2001).
19. Moreno-Pineda, E., Godfrin, C., Balestro, F., Wernsdorfer, W. & Ruben, M. Molecular

spin qudits for quantum algorithms. *Chem. Soc. Rev.* **47**, 501–513 (2018).
20. Mayländer, M., Chen, S., Lorenzo, E. R., Wasielewski, M. R. & Richert, S. Exploring Photogenerated Molecular Quartet States as Spin Qubits and Qudits. *J. Am. Chem. Soc.* **143**, 7050–7058 (2021).
21. Rozenshtein, V. *et al.* Electron Spin Polarization of Functionalized Fullerenes. Reversed Quartet Mechanism. *J. Phys. Chem. A* **109**, 11144–11154 (2005).
22. Peng, Q., Obolda, A., Zhang, M. & Li, F. Organic light-emitting diodes using a neutral π radical as emitter: The emission from a doublet. *Angew. Chemie - Int. Ed.* **54**, 7091–7095 (2015).
23. Matsuoka, R., Mizuno, A., Mibu, T. & Kusamoto, T. Luminescence of doublet molecular systems. *Coord. Chem. Rev.* **467**, 214616 (2022).
24. Li, X., Wang, Y.-L., Chen, C., Ren, Y.-Y. & Han, Y.-F. A platform for blue-luminescent carbon-centered radicals. *Nat. Commun.* **13**, 5367 (2022).
25. Ai, X. *et al.* Efficient radical-based light-emitting diodes with doublet emission. *Nature* **563**, 536–540 (2018).
26. Gamero, V. *et al.* [4-(N-Carbazolyl)-2,6-dichlorophenyl]bis(2,4,6-trichlorophenyl)methyl radical an efficient red light-emitting paramagnetic molecule. *Tetrahedron Lett.* **47**, 2305–2309 (2006).
27. Khan, Z. H. Electronic Spectra of Radical Cations and their Correlation with Photoelectron Spectra. VI. A Reinvestigation of Two-, Three-, and Four-Ring Condensed Aromatics. *Acta Phys. Pol. A* **82**, 937–955 (1992).
28. Padhye, M. R., Mcglynn, S. P. & Kasha, M. Lowest triplet state of anthracene. *J. Chem. Phys.* **23**, 588–594 (1956).
29. Biskup, T. Structure-function relationship of organic semiconductors: Detailed insights from time-resolved EPR spectroscopy. *Front. Chem.* **7**, 1–22 (2019).
30. Köhler, S. D., Höfel, S. & Drescher, M. Triplet state kinetics of anthracene studied by pulsed electron paramagnetic resonance. *Mol. Phys.* **111**, 2908–2913 (2013).
31. Mizuochi, N., Ohba, Y. & Yamauchi, S. A two-dimensional EPR nutation study on excited multiplet states of fullerene linked to a nitroxide radical. *J. Phys. Chem. A* **101**, 5966–5968 (1997).
32. Moreno-Pineda, E., Martins, D. O. T. A. & Tuna, F. Molecules as qubits, qudits and quantum gates. in *Electron Paramagnetic Resonance: Volume 27* vol. 27 146–187 (The Royal Society of Chemistry, 2021).
33. El-Sayed, M. A. Triplet state. Its radiative and nonradiative properties. *Acc. Chem. Res.* **1**, 8–16 (1968).
34. Chen, M. *et al.* Singlet Fission in Covalent Terrylenediimide Dimers: Probing the Nature of the Multiexciton State Using Femtosecond Mid-Infrared Spectroscopy. *J. Am. Chem. Soc.* **140**, 9184–9192 (2018).
35. Grüne, J., Dyakonov, V. & Sperlich, A. Detecting triplet states in opto-electronic and photovoltaic materials and devices by transient optically detected magnetic resonance. *Mater. Horizons* **8**, 2569–2575 (2021).
36. Abe, M. Diradicals. *Chem. Rev.* **113**, 7011–7088 (2013).
37. Wolfowicz, G. *et al.* Quantum guidelines for solid-state spin defects. *Nat. Rev. Mater.* **6**, 906–925 (2021).
38. Stoll, S. & Schweiger, A. EasySpin, a comprehensive software package for spectral simulation and analysis in EPR. *J. Magn. Reson.* **178**, 42–55 (2006).
39. Tait, C. E., Neuhaus, P., Anderson, H. L. & Timmel, C. R. Triplet State Delocalization in

a Conjugated Porphyrin Dimer Probed by Transient Electron Paramagnetic Resonance Techniques. *J. Am. Chem. Soc.* **137**, 6670–6679 (2015).
40. Frisch, M. J. Gaussian 16, Revision C.01. (2016).
41. Roos, B. O. The Complete Active Space Self-Consistent Field Method and its Applications in Electronic Structure Calculations. in *Advances in Chemical Physics* 399–445 (John Wiley & Sons, Ltd, 2007).
42. Angeli, C., Cimiraglia, R. & Malrieu, J.-P. N-electron valence state perturbation theory: a fast implementation of the strongly contracted variant. *Chem. Phys. Lett.* **350**, 297–305 (2001).
43. Neese, F., Wennmohs, F., Becker, U. & Riplinger, C. The ORCA quantum chemistry program package. *J. Chem. Phys.* **152**, 224108 (2020).
44. Casanova, D. & Head-Gordon, M. Restricted active space spin-flip configuration interaction approach: theory, implementation and examples. *Phys. Chem. Chem. Phys.* **11**, 9779–9790 (2009).
45. Epifanovsky, E. *et al.* Software for the frontiers of quantum chemistry: An overview of developments in the Q-Chem 5 package. *J. Chem. Phys.* **155**, 84801 (2021).

## Methods

Transient absorption spectroscopy

Transient absorption experiments were conducted on a setup pumped by a regenerative Ti:sapphire amplifier (Solstice Ace, Spectra-Physics) emitting sub-100 fs pulses centered at 800 nm at a rate of 1 kHz with a total output of 7 W. Depending on the probed spectral range and timescales, different combinations of optical systems were used.

To collect sub-ns dynamics in the visible range, frequency doubled output of the amplifier was used to seed a homebuilt broadband non-collinear optical parametric amplifier (NOPA) tuned to output 530 - 750 nm pulses with a beta barium borate (BBO) mixing crystal (Eksma Optics). Alternatively, to probe the infrared (IR) range, the output of the amplifier was used to seed a homebuilt NOPA tuned to output 1250 – 1700 nm pulses with a periodically poled stoichiometric lithium tantalate (PPSLT) mixing crystal. After chirp-correction, the white light output was split on a 50-50 beam splitter, focused to < 200 μm and used as the probe and reference beams. Wavelength-tunable pump pulses were generated in a home-built visible narrowband NOPA. The pump and probe beams were spatially overlapped at the focal point using a beam profiler, with the pump spot at least 5 times larger in diameter than the probe. Time resolution was achieved by introducing a stepped optical delay (Thorlabs DDS300-E/M) between the pump and probe pulses with a computer-controlled delay stage allowing for maximum delay of 1.9 ns, with beam wander of the probe due to changing beam pointing minimized to below 5 μm using a beam profiler. The pump pulses were chopped at 500 Hz to enable shot-to-shot referencing, which accounted for intensity fluctuations in the amplifier. After passing through the sample, the probe and reference beams were dispersed with a grating spectrometer (Shamrock SR-303i, Andor Technology) and simultaneously measured with charge-coupled device (CCD) detector arrays (Entwicklungsbüro Stresing).

To collect sub-ns dynamics in the ultraviolet range, the output of the amplifier was used to seed a homebuilt broadband NOPA tuned to output 350 - 650 nm pulses generated by focusing the 800 nm fundamental beam onto a $CaF_2$ crystal (Eksma Optics, 5 mm) connected to a digital motion controller (Mercury C-863 DC Motor Controller) after passing through a mechanical delay stage. The transmitted pulses were collected with a single line scan camera (JAI SW-4000M-PMCL) after passing through a spectrograph (Andor Shamrock SR-163).

Transient photoluminescence spectroscopy

Time-resolved PL spectra were collected using an electrically-gated intensified CCD (ICCD) camera (AndoriStar DH740 CCI-010) coupled with an image identifier tube after passing through a calibrated grating spectrometer (Andor SR303i). The spectrometer input slit width was 200 μm. The samples were excited using pump pulses obtained from a home-built narrowband NOPA driven by the same amplifier as the TA setups. A suitable long-pass filter was placed directly in front of the spectrometer to avoid the scattered laser signals entering the camera. The kinetics were obtained by setting the gate delay steps with respect to the excitation pulse. The gate widths of the ICCD were 5 ns, 50 ns, 500 ns, 5 μs and 50 μs, with overlapping time regions used to compose the decays.

Temperature-dependent measurements were performed using a closed-circuit pressurized helium cryostat (Optistat Dry BL4, Oxford Instruments), compressor (HC-4E2, Sumitomo) and temperature controller (Mercury iTC, Oxford Instruments). The vacuum level inside the cryostat was $< 10^{-5}$ mbar.

Electron spin resonance

X-band ESR was acquired with a Bruker Biospin E680 or E580 EleXSys spectrometer using a Bruker ER4118-MD5-W1 dielectric $TE01_{\delta}$ mode resonator (~9.70 GHz) in an Oxford Instruments CF935 cryostat. Q-band ESR employed an ER5106QT-2w resonator and a conventional 1.5 T electromagnet like X-band frequencies. The amplifiers for pulsed ESR, Applied Systems Engineering (ASE), had saturated powers of 1.5 kW at X-band and 180 W at Q-band. Temperature was maintained with an ITC-503S temperature controller and a CF-935SW helium flow cryostat (both Oxford Instruments). Temperature control was achieved with liquid nitrogen or helium flow and an Oxford Instruments ITC-503s temperature controller.

For laser-induced transient signals, photoexcitation was provided by Ekspla NT230 operating at a repetition rate of 50 Hz. Laser pulse energies used were 0.5 – 1 mJ, pulse lengths of 3 ns transmitted at ca. 40% through the cryostat, microwave shield and resonator windows. A liquid-crystal depolarizer (DPP-25, ThorLabs) was placed in the laser path for all measurements unless indicated. Triggering of the LASER and ESR spectrometer involved synchronisation with a Stanford Research Systems delay generator, DG645. Quadrature mixer detection was used in pulsed and continuous wave detection.

Transient cw ESR spectra were simulated using EasySpin (SI Section 2b).[38] To account for the effective deviation from isotropic ordering due to magnetophotoselection effects,[39] we introduce an ordering term of the form:

$I(\phi,\theta) = \exp(0.5*O_\theta*(3*\cos^2(\theta) -1) + O_\phi*(\sin^2(\theta)\cos(2\phi)))$, where $O_\theta$ and $O_\phi$ are the theta and phi angle ordering parameters respectively. $O_\theta$ was set to zero in all simulations.

Optically detected magnetic resonance

ODMR experiments were carried out with a modified X-band spectrometer (Bruker E300) equipped with a continuous-flow helium cryostat (Oxford ESR 900) and a microwave cavity (Bruker ER4104OR, ~9.43 GHz) with optical access. Optical irradiation was performed with a 532 nm continuous wave laser (Cobolt Samba CW 532 nm DPSSL) from one side opening of the cavity. PL was detected with a silicon photodiode (Hamamatsu S2281) on the opposite opening, using a 561 nm longpass filter to reject the excitation light. The PL signal was amplified by a current/voltage amplifier (Femto DHPCA-100). For cwODMR, PL was recorded by a lock-in detector (Ametek SR 7230), referenced by on-off-modulating the microwaves with a modulation frequency of 547 Hz. The microwaves were generated with a microwave signal generator (Anritsu MG3694C), amplified to 3 W (microsemi) and guided into the cavity. For trODMR, PL was recorded by a digitizer card (GaGe Razor Express 1642 CompuScope), whereby a pulse blaster card (PulseBlasterESR-PRO) triggered the digitizer card and produced microwave pulses for a set length. The microwaves were generated with the same microwave signal generator as in cwODMR, whereby they were amplified to 5 W by a traveling wave tube amplifier (TWTA, Varian VZX 6981 K1ACDK) and guided into the cavity.

Theoretical calculations

The doublet ground-state $^2[D_0S_0]$ and quartet $^4[D_0T_1]$ of the R-A monoradicals TTM-1Cz-An and TTM-1Cz-PhAn were optimised by means of the Unrestricted Kohn-Sham (UKS) formalism within the Density Functional Theory (DFT) framework, using the ωB97X-D exchange-correlation functional and the 6-31G(d,p) basis set. In both $^2[D_0S_0]$ and $^4[D_0T_1]$ optimised structures, the spin contamination is predicted to be negligible (less than 5%). The R-A-R biradical $(TTM\text{-}1Cz)_2$-An ground-state $^3[D_0S_0D_0]$ was optimised with the same level of theory as described above. Broken-Symmetry (BS) DFT calculations pointed to a degeneracy between the triplet $^3[D_0S_0D_0]$ and the BS singlet ground-state $^1[D_0S_0D_0]$, which was found to lie less than 0.05 $cm^{-1}$ above the $^3[D_0S_0D_0]$ configuration. These calculations were performed with the Gaussian16 software.[40]

In order to gain access to all the relevant configuration state functions (CSFs) of R-A, state-averaged Complete Active Space Self-Consistent Field (CASSCF) calculations were performed on the optimised monoradical TTM-1Cz-An ground-state $^2[D_0S_0]$ structure, using the Def2-TZVP basis set.[41] On top of a converged CASSCF wave function, strongly contracted second order N-Electron Valence state Perturbation Theory (NEVPT2) calculations were performed to recover the missing dynamic electronic correlation at the CASSCF level.[42] Since the CASSCF wave function is expanded in terms of Slater determinants computed in a restricted-open shell formalism, both CASSCF and NEVPT2 methods provide excited states which are free from spin contamination. The same computational methods were applied to on the optimised biradical $(TTM\text{-}1Cz)_2$-An ground-state $^3[D_0S_0D_0]$ structure, using a smaller basis set Def2-SVP to reduce the computational costs. Such calculations were run with the ORCA 4.2 code.[43]

The electronic couplings between the diabatic doublet states $^2[D_0T_1]$, $^2[D_1S_0]$ and $^2$CT of the monoradical TTM-1Cz-An were estimated via a diabatization procedure considering the Boys localization scheme, where the *adiabatic* states were computed with the Restricted Active Space Configuration Interaction (RAS-CI) method,[44] along with the Def2-SVP basis set within the Q-Chem 5.4 software.[45] In the RAS-CI formalism the molecular orbital (MO) space is divided into three subspaces, namely RAS1, RAS2 and RAS3. The excited configurations are generated by an excitation operator ($\hat{R}$) acting on the Restricted-Open shell Hartree-Fock (ROHF) reference wave function ($\phi_0$):

$$|\Psi^{RAS-CI}\rangle = \hat{R}|\phi_0\rangle$$

In the current RAS-CI implementation of Q-Chem 5.4, $\hat{R}$ is defined as:

$$\hat{R} = \hat{r}^{RAS2} + \hat{r}^{hole} + \hat{r}^{particles}$$

where: $\hat{r}^{RAS2}$ generates all the possible electronic configurations (singles, doubles, triples, etc.) in the RAS2 subspace, corresponding to a full CI treatment within the selected subspace ; $\hat{r}^{hole}$ generates electronic configurations by promoting single excitations from the RAS1 to RAS2, creating *n* holes in the RAS1 subspace; analogously, $\hat{r}^{particles}$ generates electronic configurations by promoting electrons from RAS2 to RAS3, thus creating *m* particles in the RAS3. In our case, the RAS2 is built from 11 electrons in 10 orbitals which allows to recover all the relevant states (*i.e.,* $^2[D_1S_0]$, $^{2/4}[D_0T_1]$ and $^2$CT) predicted at the NEVPT2 level; the RAS1 (RAS3) allows for the creation of 6 holes (particles), while the remaining MOs remain doubly occupied (unoccupied). Alike CASSCF and NEVPT2, the RAS-CI approach allows to obtain excited states free from spin-contamination.

In the Boys localization scheme, the diabatic states are written as a linear combination of *adiabatic* states:

$$|\Xi_\Gamma\rangle = \sum_{J}^{N_{Adiab}} |\Psi_J\rangle U_{J\Gamma}$$

with $|\Xi_\Gamma\rangle$ the Γ-th diabatic state, $|\Psi_J\rangle$ the J-th *adiabatic* state and $U_{J\Gamma}$ the element of the rotation matrix from the *adiabatic* to the diabatic representation. With such a scheme, the electric dipole moment difference between each pair of diabatic states is maximised:

$$f_{Boys}(\boldsymbol{U}) = f_{Boys}(\{\Xi\}) = \sum_{\Gamma,\Delta=1}^{N_{Adiab}} |\langle\Xi_\Gamma|\hat{\mu}|\Xi_\Gamma\rangle - \langle\Xi_\Delta|\hat{\mu}|\Xi_\Delta\rangle|^2$$

As a result, we obtain the rotation matrix **U,** which transforms the Hamiltonian from a (diagonal) *adiabatic* to a (non-diagonal) diabatic representation. The matrix elements of the diabatic Hamiltonian $\langle\Xi_\Gamma|\hat{H}|\Xi_\Delta\rangle$ represent either the diabatic state energy when $\Gamma = \Delta$ or the electronic coupling between diabatic states when $\Gamma \neq \Delta$. The diabatization was carried out on top of the RAS-CI excited states by systematically increasing the size of the diabatization, *i.e.,* increasing the number of the *adiabatic* states. In the 3x3 diabatization only the *adiabatic* state associated with $^2[D_1S_0]$, $^2[D_0T_1]$ and $^2$CT were considered, while for the 9x9 and 17x17 all the first 9 and 17 *adiabatic* excited states were introduced, respectively.

**Acknowledgments:** We thank Leah Weiss and Claudia Tait for useful discussions. This work was supported by the following funding sources: European Research Council (ERC), European Union's Horizon 2020 research and innovation programme grant agreement no. 101020167 (S.G., R.H.F.); Engineering and Physical Sciences Research Council (EPSRC) Cambridge NanoDTC, EP/S022953/1 (S.G., B.D.); Simons Foundation, grant no. 601946 (R.H.F.); Royal Society, University Research Fellowship, grant no. URF\R1\201300 (E.W.E.); National Natural Science Foundation of China, grant no. 51925303 (K.L., F.L.); Deutsche Forschungsgemeinschaft (DFG), Research Training School "Molecular biradicals: Structure, properties and reactivity" (GRK2112) (J.G., A.S.); Bavarian Ministry of the Environment and Consumer Protection, the Bavarian Network "Solar Technologies Go Hybrid" (J.G., V.D.); John Fell OUP Research Fund, grant no.: 0007019 (W.K.M.); EPSRC, grant nos: EP/V036408/1 and EP/L011972/1 (W.K.M.); Department of Chemistry, University of Oxford (W.K.M.); Fonds de la Recherche Scientifiques de Belgique (F.R.S.-FNRS), grant no. 2.5020.11 (D.B., Y.O., G.L., G.R., D.V.); Walloon Region, grant no. 1117545 (D.B., Y.O., G.L., G.R., D.V.); Fonds pour la formation à la Recherche dans l'Industrie et dans l'Agriculture of the F.R.S.-F.N.R.S. (G.R.); Fonds de la Recherche Scientifique-FNRS, grant no. F.4534.21 (MIS-IMAGINE) (Y.O.); Ministerio de Ciencia e Innovación (MICINN) of Spain, project PID2019-109555GB-I00 (C.T., D.C.); Eusko Jaurlaritza/Basque Government, projects PIBA19-0004 and 2019-CIEN-000092-01 (C.T., D.C.). Computational resources were provided by the Consortium des Équipements de Calcul Intensif (CÉCI) and the Tier-1 supercomputer of the Fedération Wallonie-Bruxelles; D.B. is a FNRS Research Director. C. T. is supported by Donostia International Physics Centre (DIPC) and Gipuzkoa's council joint program Women and Science. C.T. and D.C. are thankful for the technical and human support provided by the DIPC Computer Centre.

**Author Contributions:** S.G. and B.H.D. performed the photophysical measurements. S.G. and W.K.M. performed the ESR measurements. J.G. performed the ODMR measurements. K.L. synthesized the compounds and performed the chemical characterisation. G.L., G.R., D.V., C.T., D.C., D.B. and Y.O. developed the computational methods and performed the calculations. P.M. performed additional chemical characterisation. A.S.R. performed the frozen solution 77 K optical measurements. V.D., A.S., D.C., D.B., Y.O., F.L. and R.H.F. supervised their group members involved in the project. S.G., R.H.F. and E.W.E. designed the experiments and analysed the data. R.H.F. and E.W.E. conceived the project. S.G., R.H.F. and E.W.E. wrote the manuscript with input from all authors.

**Additional Information:** Correspondence and requests for materials should be addressed to S.G. (sg911@cam.ac.uk), R.H.F. (rhf10@cam.ac.uk) and E.W.E. (emrys.evans@swansea.ac.uk). Supplementary Information is available for this paper. Authors declare that they have no competing interests. The data supporting this study are available at the University of Cambridge Repository [URL to be added].

**Extended Data Figures:**

1. Transient optical absorption of TTM-1Cz-An
2. Steady state photophysics
3. Temperature dependent transient photoluminescence
4. Ground state ESR
5. Transient ESR
6. ODMR
7. Room temperature pulsed ESR

**Extended Data Table:**

1. Spin Hamiltonian modelling

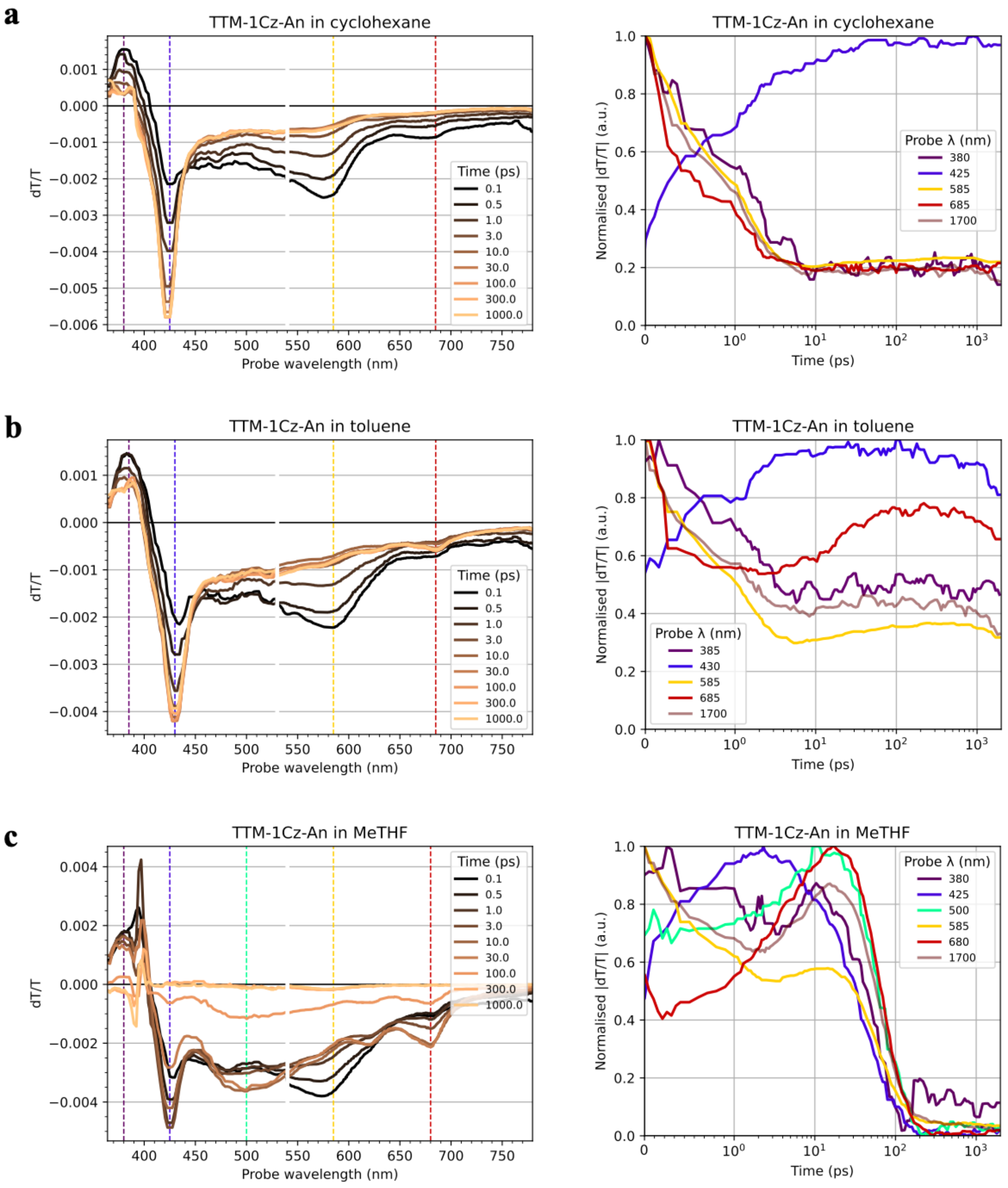

**Extended Data Fig. 1. TTM-1Cz-An transient optical absorption.** Spectral (left) and kinetic (right) TA slices of 200 μM solutions in **a,** cyclohexane, **b,** toluene, and **c,** 2-methyltetrohydrofuran (MeTHF). In kinetics, time axis scale is linear until 1 ps, and logarithmic thereafter; and absolute values of the TA signal normalised to their peak are shown. Excitation at 600 nm with fluences of 40 μJ $cm^{-2}$, matched between 360-580 and 530-780 nm probe region experiments for each plot. Solvatochromism observed after ca. 5 ps, i.e. following rapid energy transfer. CT signature at 685 nm absent in cyclohexane (least polar). In MeTHF (most polar) non-radiative decay occurs after the emergence of the CT band.

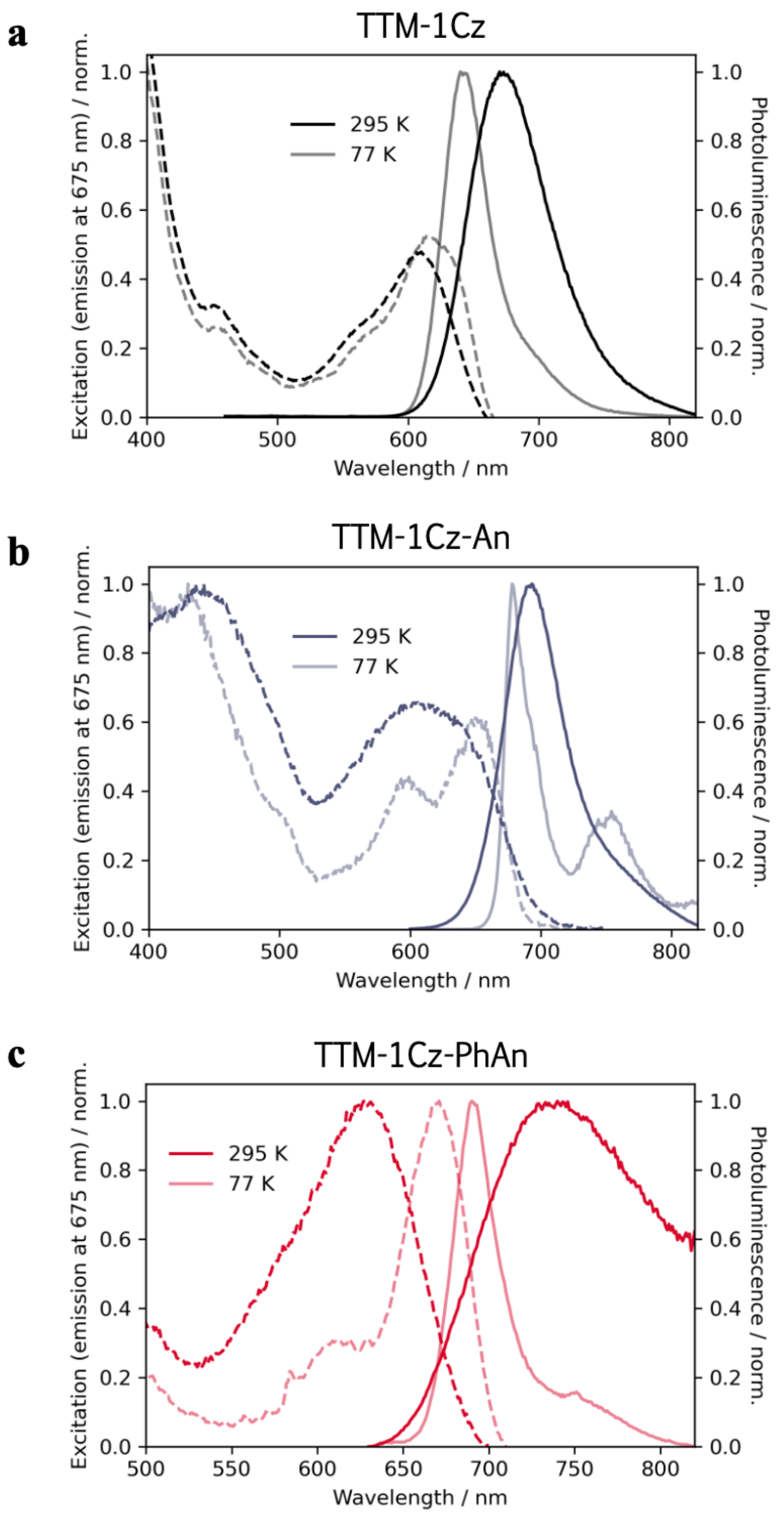

**Extended Data Fig. 2. Steady state photophysics.** Excitation and emission scans at 295 and 77 K in 200 μM toluene solutions of **a,** TTM-1Cz, **b,** TTM-1Cz-An, and **c,** TTM-1Cz-PhAn. Vibronic progression reminiscent of pure anthracene is seen in TTM-1Cz-An and TTM-1Cz-PhAn at low temperatures.

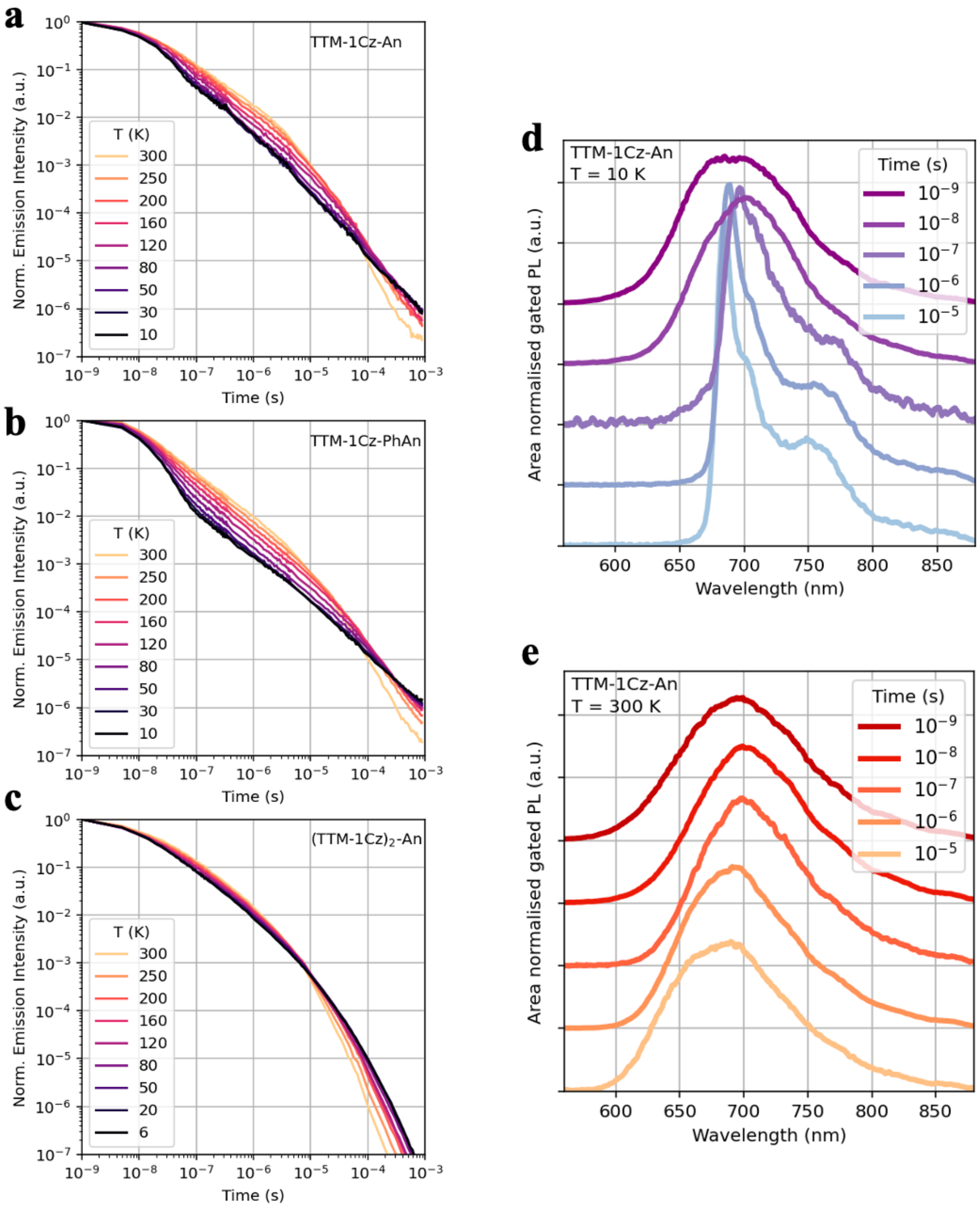

**Extended Data Fig. 3. Temperature dependent transient photoluminescence.** Normalised emission kinetics in 5% PMMA films of **a,** TTM-1Cz-An ,**b,** TTM-1Cz-PhAn, and **c,** (TTM-1Cz)$_2$-An. All materials show temperature-activated delayed emission. **d,** Time gated emission spectra in 5% TTM-1Cz-An in PMMA films at 10 K, and **e,** 300 K. All measurements performed in vacuum after excitation at 520 nm with fluence of 26 μJ cm$^{-2}$ (for R-A) or 530 nm with fluence of 8.7 μJ cm$^{-2}$ (for R-A-R).

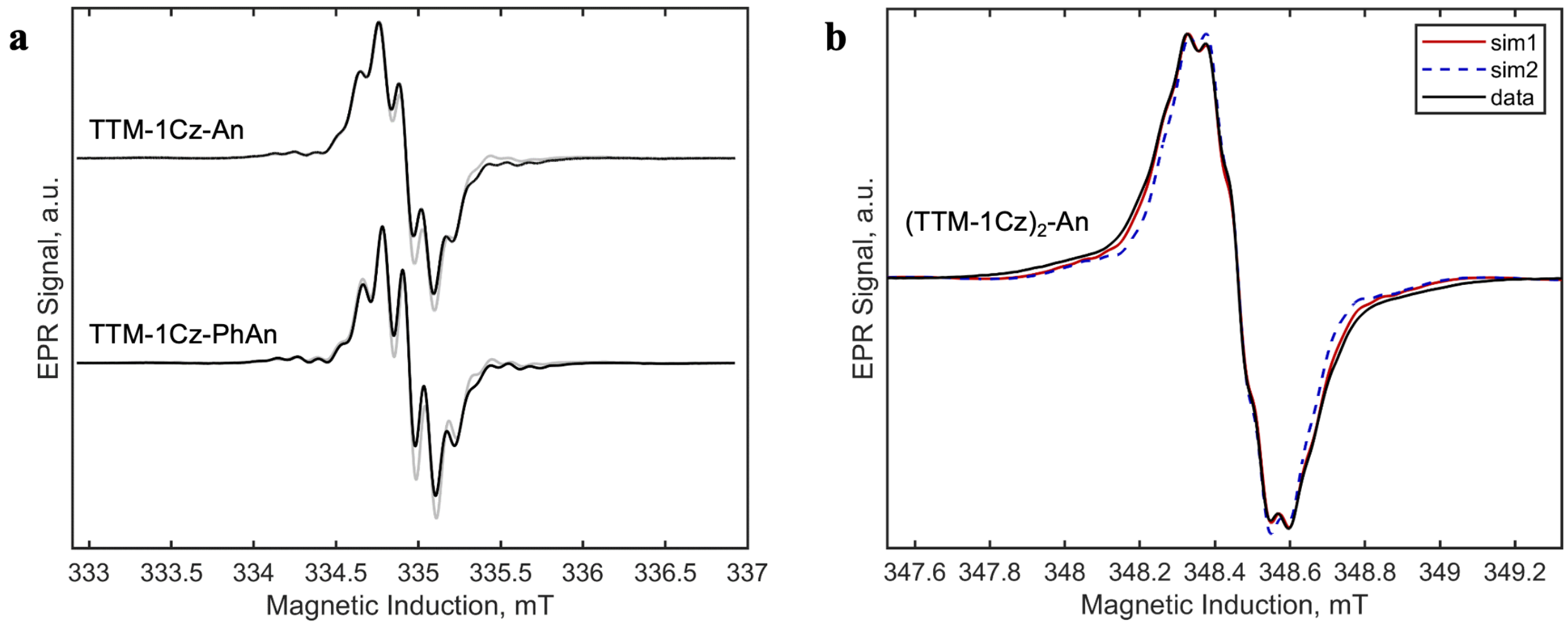

**Extended Data Fig. 4. Ground state ESR. a,** Room temperature X-band cw-ESR of 200 μM toluene solutions of TTM-1Cz-An and TTM-1Cz-PhAn in the dark confirming their spin doublet ground states. The black traces are data, and the grey lines are simulations. The respective acquisition parameter values were microwave frequencies 9.3923 & 9.3927 GHz, field modulations of 5 & 20 μT, microwave powers of 200 & 100 nW, both being non-saturating. Simulation employed experimental parameter values with S = 1/2, g = 2.00355 (±0.00005), six $A_{\text{iso}}(^1\text{H}) = 0.121$, one $A_{\text{iso}}(\text{C}_{\text{nat. abund.}}) = 2.7$ mT, and six $A_{\text{iso}}(\text{C}_{\text{nat. abund.}}) = 11$ mT. **b,** cw-ESR at X-band of (TTM-1Cz)$_2$-An in toluene at a concentration of 200 μM, collected at 295 K and shown in the black trace with two simulations. The acquisition microwave frequency was 9.7715 GHz with power of 2 μW, the field modulation amplitude was 2 μT. The best fit here is provided by increasing the number of contributing protons with $A(^1\text{H}) = 0.6$ MHz from twice the TTM-1Cz-An simulation in panel a (12 protons, sim2) to 14 protons (sim1). As $J > A$, the hyperfine values are divided by a factor of two compared to TTM-1Cz-An. The conformational degrees of freedom in solution provide a range of exchange values, leading to an average in excess of the apparent hyperfine values.

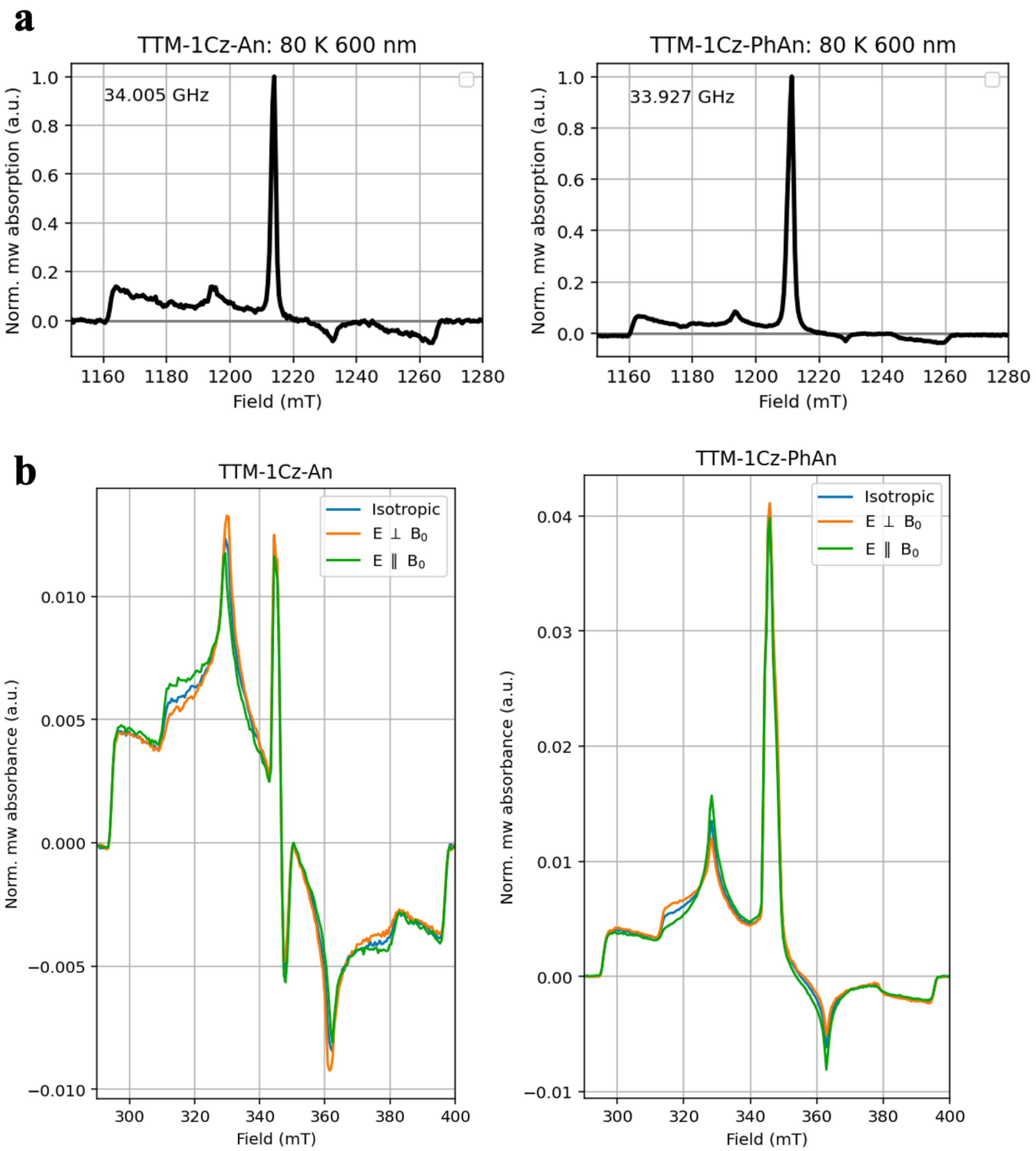

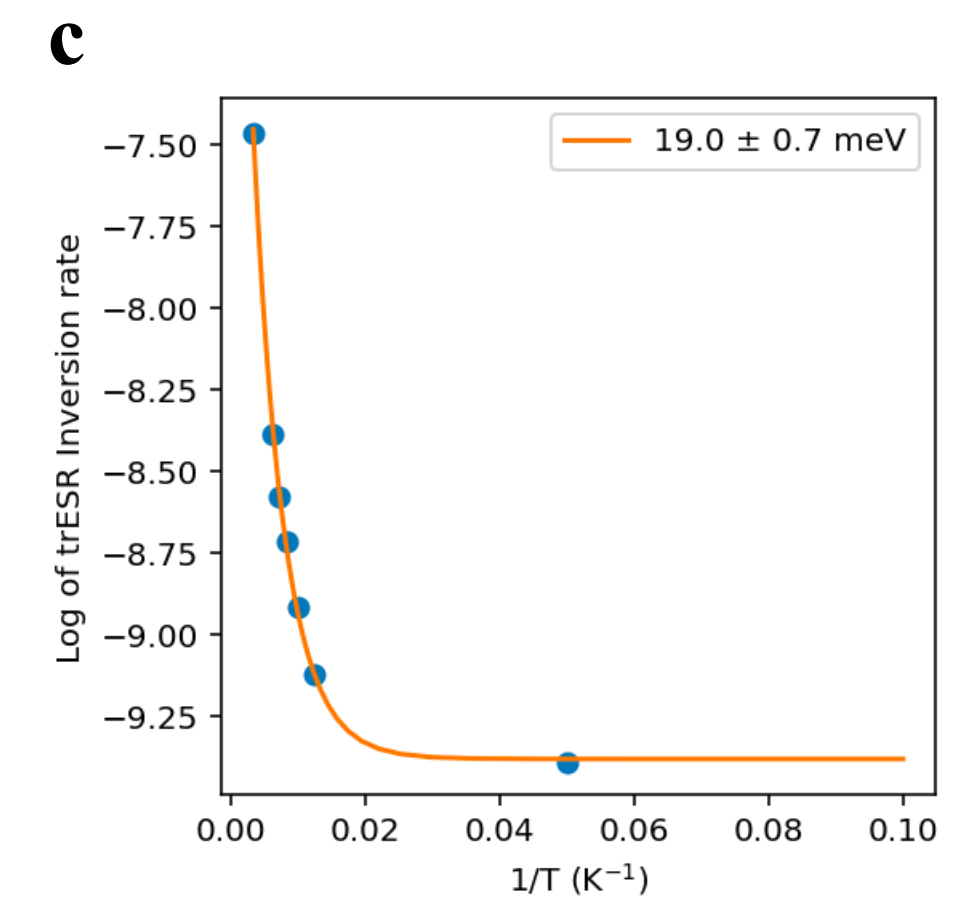

**Extended Data Fig. 5. R-A trESR. a,** Q-band FF trESR on frozen 200 μM toluene solutions, 1-4 μs after excitation with 0.4 mJ 600 nm pulses. The lineshape is analogous to the X-band FF spectra, with no level crossings detected at fields up to 1.3 T. **b,** Magnetophotoselection in X-band FF trESR**.** Dilute toluene solutions at 80 K following 0.7 mJ 600 nm excitation at 40 dB microwave attenuation. Signals normalized by total area. No selection observed for z transitions, and opposite dependence for TTM-1Cz-An and TTM-1Cz-PhAn seen for x/y transitions. **c,** Arrhenius plot of quartet polarisation inversion dynamics at in TTM-1Cz-An.

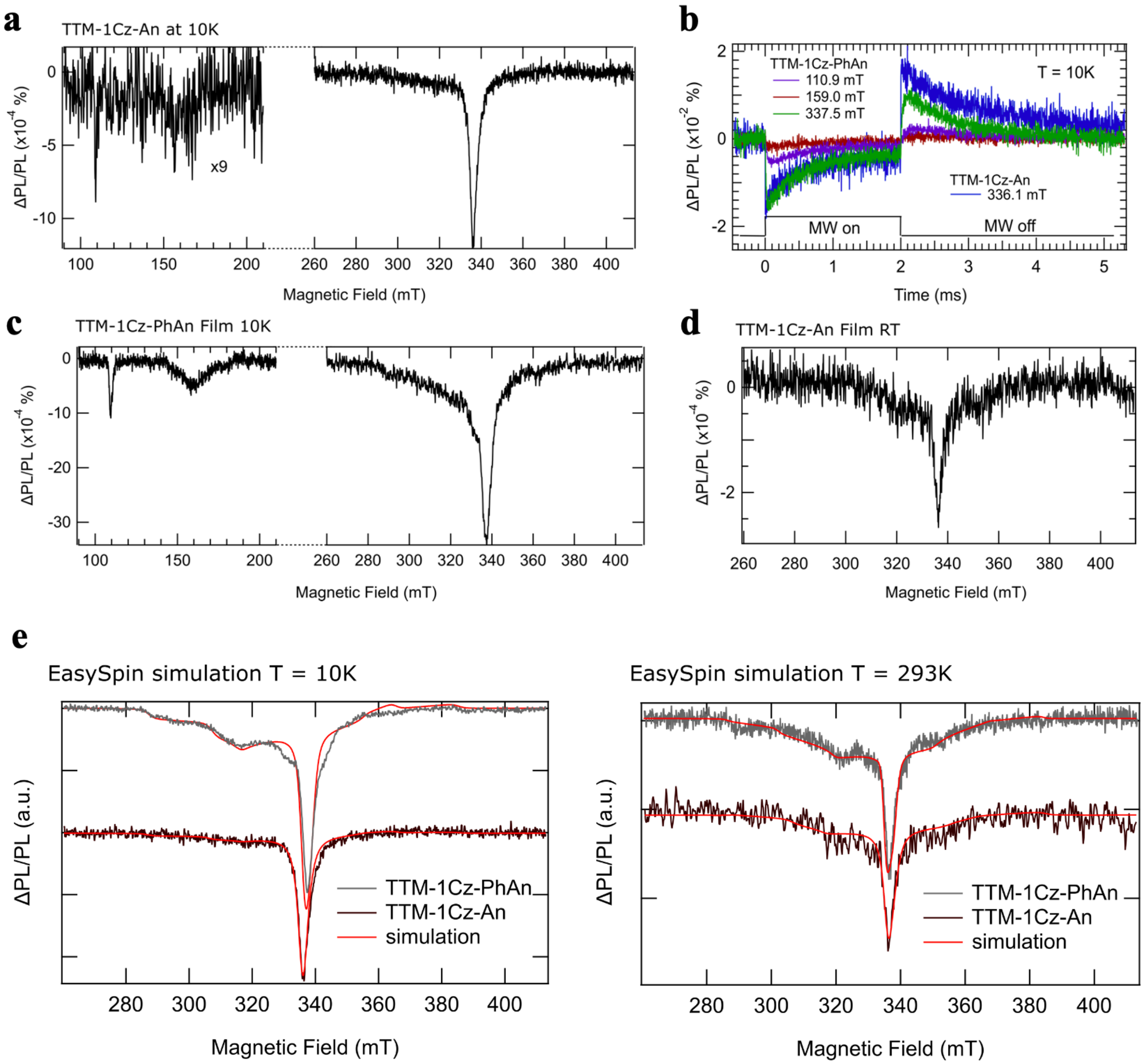

**Extended Data Fig. 6. ODMR. a,** cwODMR at 10K of dilute frozen toluene solution of TTM-1Cz-An. **b,** trODMR at 10 K for TTM-1Cz-PhAn HF and FF signals (purple, red, green) and TTM-1Cz-An FF signal (blue), revealing a negative sign for all signals. **c,** cwODMR spectra of 5% TTM-1Cz-PhAn in PMMA film at 10 K. **d,** cwODMR spectrum of 5% TTM-1Cz-An in PMMA film at 293 K, showing quartet contribution at room temperature. **e,** Simulations (red) for TTM-1Cz-PhAn (grey) and TTM-1Cz-An (black) at T = 10 K (left) and T =293 K (right).

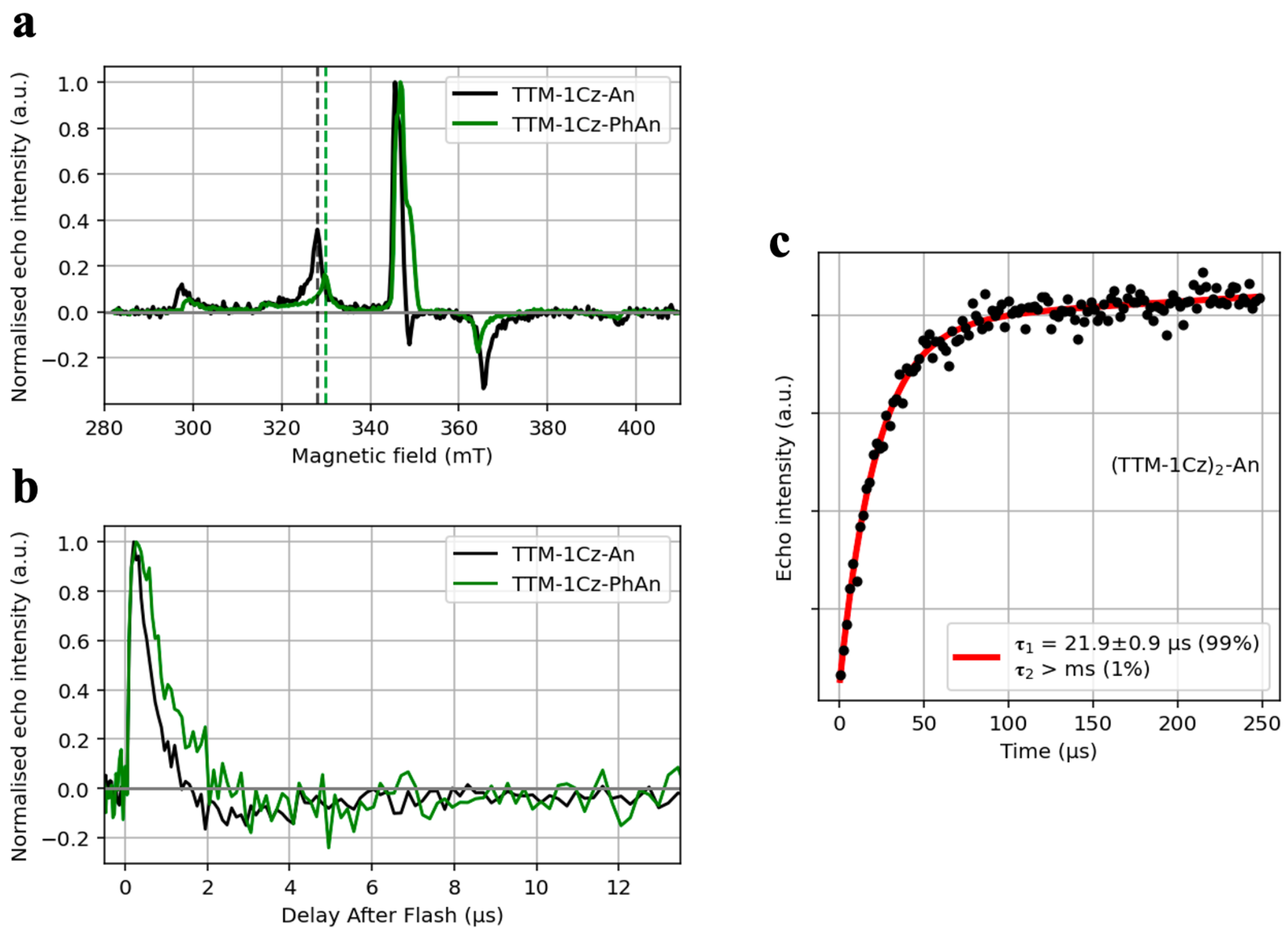

**Extended Data Fig. 7. Room temperature pulsed ESR. a,** Echo Detected Field Sweep (EDFS) of excited quartet states of R-A at 295 K in 5% radical in PMMA 0.5 μs after 1.4 mJ 600 nm excitation. **b,** Delay After Flash (DAF) scans at quartet field positions (marked with dashed vertical lines in panel a). **c,** Spin-lattice relaxation of R-A-R ground state. 5% $(TTM\text{-}1Cz)_2$-An in PMMA film at 295 K. Experiment performed at field corresponding to $g = 2.00355$ at X-band (345.65 mT) without light excitation. Inversion recovery pulse sequence.

**a**

| Sample | $\lvert D_Q \rvert$, $\lvert E_Q \rvert$ (MHz) | $O_\phi$ | time (µs) | Populations $Q_{-1/2}$, $Q_{+1/2}$, $Q_{-3/2}$, $Q_{+3/2}$, $D_{-1/2}$, $D_{+1/2}$ |
|---|---|---|---|---|
| TTM-1Cz-An 80 K, 200 µM | 705, 79 | -1.3 | 1 | 0.42, 0.35, 0.10, 0.13, 0.00, 0.10 |
| | | | 15 | 0.10, 0.02, 0.43, 0.45, 0.00, 0.08 |
| TTM-1Cz-An 150 K, 200 µM | 711, 90 | -1.4 | 1 | 0.41, 0.41, 0.07, 0.11, 0.00, 0.07 |
| | | | 9 | 0.00, 0.01, 0.51, 0.48, 0.00, 0.09 |
| TTM-1Cz-An 295 K, PMMA | 702, 59 | -2.1 | 1 | 0.45, 0.40, 0.00, 0.15, 0.00, 0.10 |
| | | | 3 | 0.14, 0.18, 0.32, 0.36, 0.00, 0.07 |
| TTM-1Cz-PhAn 80 K, 200 µM | 687, 64 | -1.7 | 1 | 0.36, 0.30, 0.08, 0.26, 0.00, 0.15 |
| | | | 11 * | 0.42, 0.37, 0.00, 0.21, 0.00, 0.12 |
| TTM-1Cz-PhAn 150 K, 200 µM | 694, 73 | -1.8 | 1 | 0.33, 0.30, 0.13, 0.23, 0.00, 0.10 |
| | | | 11 | 0.14, 0.24, 0.33, 0.29, 0.05, 0.00 |
| TTM-1Cz-PhAn 295 K, PMMA | 690, 70 | -1.7 | 1 | 0.38, 0.27, 0.04, 0.31, 0.00, 0.19 |
| | | | 5 | 0.21, 0.24, 0.27, 0.27, 0.00, 0.15 |

**b**

| Material | Temp. (K) | $\lvert D_Q \rvert$, $\lvert E_Q \rvert$ (MHz) | Lw (mT) |
|---|---|---|---|
| TTM-1Cz-An | 10 | 690, (50)* | 1.5, 3.0 |
| TTM-1Cz-Ph-An | 10 | 690, 50 | 2.0, 2.0 |
| TTM-1Cz-An | 293 | 690, (50)* | 1.5, 3.0 |
| TTM-1Cz-Ph-An | 293 | 690, 80 | 1.5, 1.5 |

**Extended Data Table 1. R-A Spin Hamiltonian modelling. a,** Simulation parameters for full-field X-band trESR spectra. $g_T$ = 2.003, $g_R$ = 2.00355. Linewidth was set to 1.5 mT in all simulations. Populations at maximum absolute polarisation times before and after inversion. *: Not inverted within acquisition timescale. **b,** Simulation parameters for ODMR spectra. $g_T$ = 2.003, $g_R$ = 2.003, $O_\phi$ = 0. Linewidth (Lw) given in Gaussian, Lorentzian. *: E value cannot be determined given the too weakly pronounced X and Y transitions.